\documentclass[
psamsfonts,
intlimits,
sumlimits,
namelimits,
12dd,
a4paper,
pdftex,
]{kgeneric}

\usepackage{arxiv}

\newif\iftexht\texhtfalse
\makeatletter
\@ifpackageloaded{tex4ht}{\texhttrue}{\texhtfalse}
\makeatother

\IfFileExists{ifpdf.sty}{\usepackage{ifpdf}}{}
\IfFileExists{ifxetex.sty}{\usepackage{ifxetex}}{}
\IfFileExists{ifluatex.sty}{\usepackage{ifluatex}}{}

\usepackage{xifthen}

\usepackage{cmap}
\usepackage{mathtext}

\usepackage[utf8]{inputenc}
\usepackage[T2A]{fontenc}
\usepackage[english,frenchb,german,russian]{babel}

\makeatletter
\@ifpackageloaded{ucs}{%
	\PrerenderUnicode{АБВГДЕЁЖЗИЙКЛМНОПРСТУФХЦЧШЩЬЫЪЭЮЯабвгдеёжзийклмнопрстуфхцчшщьыъэюя–}
}
\makeatother

		\def\copyright{\textcopyright}

\RequirePackage{scrlfile}
\BeforePackage{bm}{%

}

\ifpdf
	\usepackage{microtype}
	\UseMicrotypeSet[protrusion]{alltext}
\fi

\usepackage{etex}
\usepackage{etextools}

\usepackage{fixltx2e}
\usepackage[shortcuts]{extdash}

\usepackage{setspace}
\usepackage[perpage,bottom,multiple,stable]{footmisc}

\DeclareUnicodeCharacter{2212}{-}

\RequirePackage[thmmarks]{ntheorem}
\AtBeginDocument{
	\captionsetup[table]{format=plain,position=above,justification=centering,labelsep=period,singlelinecheck=false}
}

\makeatletter
\long\def\parseauthor#1 #2 #3\nil%
{%
	\edef\@tempa{#1}
	\ifx\@tempa\@empty
		\edef\authorinitials{#2}
		\edef\authorinitialsplain{#2}
		\edef\authorfamily{#3}
	\else
		\def\authorinitials{#2\leavevmode\nobreak\thinspace#3}
		\edef\authorinitialsplain{#2 #3}
		\def\authorfamily{#1}
	\fi
}
\makeatother

\makeatletter
\def\author@fmt#1#2#3#4%
{%

	\@newelemtrue
	\if@firstauthor
		\first@author \global\@firstauthorfalse \fi
	\ifnum\prev@elem=\e@author \global\@newelemfalse \fi

	\parseauthor#3 \nil

	\protected@edef\@tempd{#4}
	\ifx\@tempd\@empty
		\protected@edef\authorFamIn{\authorfamily\kern 0.3em \authorinitials}
		\protected@edef\authorInFam{\authorinitials\kern 0.3em \authorfamily}
		\protected@xdef\authorFamily{\authorfamily\xspace}
		\protected@xdef\authorInitials{\authorinitials\xspace}
		\protected@xdef\authorInitialsPlain{\authorinitialsplain\xspace}
	\else
		\protected@edef\authorFamIn{\protect\csname\@tempd \endcsname{\authorfamily\kern 0.3em \authorinitials}}
		\protected@edef\authorInFam{\protect\csname\@tempd \endcsname{\authorinitials\kern 0.3em \authorfamily}}
	\fi

	\ifx\authors\@empty
	\protected@xdef\authorstoc{\authorFamIn}%
	\protected@xdef\authorsrunning{\authorFamIn}%

	\protected@edef\@tempb{#2}
	\ifx\@tempb\@empty
		\ifx\@tempd\@empty
			\g@addto@macro\authors{\protect\mbox{#3}}%
		\else
			\g@addto@macro\authors{\protect\csname\@tempd \endcsname{#3}}%
		\fi
		\else
			\ifx\@tempd\@empty
				\g@addto@macro\authors{\protect\mbox{#3}$^{\mathrm{{#2}}}$}%
			\else
				\g@addto@macro\authors{\protect\csname\@tempd \endcsname{#3}$^{\mathrm{{#2}}}$}%
			\fi
		\fi
	\else
		\protected@xdef\authorstoc{\authorstoc \and \authorFamIn}%
		\ifnum0\n@author@ > \PFUNumRunningauthor
			\ifnum0#1 = \PFUNumRunningauthor 
				\g@addto@macro\authorsrunning{ \etal}%
			\fi
		\else
			\protected@xdef\authorsrunning{\authorsrunning \and \authorFamIn}
		\fi
		\protected@edef\@tempb{#2}
		\ifx\@tempb\@empty
			\ifx\@tempd\@empty
				\g@addto@macro\authors{\and \protect\mbox{#3}}%
			\else
				\g@addto@macro\authors{\and \protect\csname\@tempd \endcsname{#3}}%
			\fi
		\else
			\ifx\@tempd\@empty
				\g@addto@macro\authors{\and \protect\mbox{#3}$^{\mathrm{{#2}}}$}%
			\else
				\g@addto@macro\authors{\and \protect\csname\@tempd \endcsname{#3}$^{\mathrm{{#2}}}$}%
			\fi
		\fi
	\fi
} 

\def\altauthor@fmt#1#2#3#4%
{%

	\if@firstauthor
		\first@author \global\@firstauthorfalse \fi
	\ifnum\prev@elem=\e@author \global\@newelemfalse \fi
	\parseauthor#3 \nil

	\edef\@tempd{#4}
	\ifx\@tempd\@empty
		\protected@xdef\altauthorFamIn{\authorfamily\kern 0.3em \authorinitials}
		\protected@xdef\altauthorInFam{\authorinitials\kern 0.3em \authorfamily}
		\protected@xdef\altauthorFamily{\authorfamily\xspace}
		\protected@xdef\altauthorInitials{\authorinitials\xspace}
		\protected@xdef\altauthorInitialsPlain{\authorinitialsplain\xspace}
	\else
		\protected@xdef\altauthorFamIn{\protect\csname\@tempd \endcsname{\authorfamily\kern 0.3em \authorinitials}}
		\protected@xdef\altauthorInFam{\protect\csname\@tempd \endcsname{\authorinitials\kern 0.3em \authorfamily}}
	\fi

	\ifx\@empty\altauthors
		\protected@xdef\altauthorstoc{\altauthorFamIn}%
		\protected@xdef\altauthorsrunning{\altauthorFamIn}%
		\edef\@tempb{#2}
		\ifx\@tempb\@empty
			\ifx\@tempd\@empty
				\protected@xdef\altauthors{\altauthorInFam}%
			\else
				\g@addto@macro\altauthors{\protect\csname\@tempd \endcsname{#3}}%
			\fi
		\else
			\ifx\@tempd\@empty
				\g@addto@macro\altauthors{\protect\mbox{#3}$^{\mathrm{{#2}}}$}%
			\else
				\g@addto@macro\altauthors{\protect\csname\@tempd \endcsname{#3}$^{\mathrm{{#2}}}$}%
			\fi
		\fi
	\else
		\protected@xdef\altauthorstoc{\altauthorstoc \altand \altauthorFamIn}%
		\ifnum0\n@author@ > 3
			\ifnum0#1 = 2
			\fi
		\else
			\protected@xdef\altauthorsrunning{\altauthorsrunning \altand	\altauthorFamIn}
		\fi
		\edef\@tempb{#2}
		\ifx\@tempb\@empty
			\ifx\@tempd\@empty
				\protected@xdef\altauthors{\altauthors \altand \altauthorInFam}%
			\else
				\g@addto@macro\altauthors{\altand \protect\fbox{#3}}%
			\fi
		\else
			\ifx\@tempd\@empty
				\g@addto@macro\altauthors{\altand \protect\mbox{#3}$^{\mathrm{{#2}}}$}%
			\else
				\g@addto@macro\altauthors{\altand \protect\fbox{#3}$^{\mathrm{{#2}}}$}%
			\fi
		\fi
	\fi
} 

\def\and{\unskip{}, \ignorespaces}
\def\altand{\unskip{}, \ignorespaces}

\makeatother

\makeatletter

\newcommand\backmatter{%
		\if@openright
			\cleardoublepage
		\else
			\clearpage
		\fi
	}
\makeatother

\usepackage{upgreek}

\AtBeginDocument{%
	\ifdef{\letupgreekdefault}{\relax}{\def\letupgreekdefault{0}}
	\ifdim \letupgreekdefault pt = 0pt
		\relax
	\else
		\let\alpha\upalpha
		\let\beta\upbeta
		\let\gamma\upgamma
		\let\delta\updelta
		\let\epsilon\upepsilon
		\let\zeta\upzeta
		\let\eta\upeta
		\let\theta\uptheta
		\let\iota\upiota
		\let\kappa\upkappa
		\let\lambda\uplambda
		\let\mu\upmu
		\let\nu\upnu
		\let\xi\upxi
		\let\pi\uppi
		\let\rho\uprho
		\let\sigma\upsigma
		\let\tau\uptau
		\let\upsilon\upupsilon
		\let\phi\upphi
		\let\chi\upchi
		\let\psi\uppsi
		\let\omega\upomega
		\let\varepsilon\upvarepsilon
		\let\vartheta\upvartheta
		\let\varpi\upvarpi
		\let\varrho\upvarrho
		\let\varsigma\upvarsigma
		\let\varphi\upvarphi
	\fi
	\ifdef{\letUpgreekdefault}{\relax}{\def\letUpgreekdefault{0}}
	\ifdim \letUpgreekdefault pt = 0pt
		\relax
	\else
		\let\Gamma\Upgamma
		\let\Delta\Updelta
		\let\Theta\Uptheta
		\let\Lambda\Uplambda
		\let\Xi\Upxi
		\let\Pi\Uppi
		\let\Sigma\Upsigma
		\let\Upsilon\Upupsilon
		\let\Phi\Upphi
		\let\Psi\Uppsi
		\let\Omega\Upomega
	\fi
}

\AtEndPreamble{%
	 \ifundef\becquerel{\def\becquerel{}}{}
	 \ifundef\dptr{\def\dptr{}}{}
	 \ifundef\degreeCelsius{\def\degreeCelsius{}}{}
	 \ifundef\celsius{\def\celsius{}}{}
	 \ifundef\dyn{\def\dyn{}}{}
	 \ifundef\atm{\def\atm{}}{}
	 \ifundef\erg{\def\erg{}}{}
	 \ifundef\earthmass{\def\earthmass{}}{}
	 \ifundef\terramass{\def\terramass{}}{}
	 \ifundef\solarmass{\def\solarmass{}}{}
	 \ifundef\solmass{\def\solmass{}}{}
	 \ifundef\earthradius{\def\earthradius{}}{}
	 \ifundef\terraradius{\def\terraradius{}}{}
	 \ifundef\solarradius{\def\solarradius{}}{}
	 \ifundef\solradius{\def\solradius{}}{}
	 \ifundef\solarluminosity{\def\solarluminosity{}}{}
	 \ifundef\solluminosity{\def\solluminosity{}}{}
	 \ifundef\solarirradiance{\def\solarirradiance{}}{}
	 \ifundef\sollirradiance{\def\sollirradiance{}}{}
	 \ifundef\units{\def\units{}}{}
	 \ifundef\piece{\def\piece{}}{}
	 \ifundef\pieces{\def\pieces{}}{}
	 \ifundef\items{\def\items{}}{}
	 \ifundef\circulation{\def\circulation{}}{}
	 \ifundef\copies{\def\copies{}}{}
	 \ifundef\percent{\def\percent{}}{}
	 \ifundef\procent{\def\procent{}}{}
	 \ifundef\permille{\def\permille{}}{}
	 \ifundef\promille{\def\promille{}}{}
	 \ifundef\permyriad{\def\permyriad{}}{}
	 \ifundef\promyriad{\def\promyriad{}}{}
	 \ifundef\crystalmass{\def\crystalmass{}}{}
	 \ifundef\absorbance{\def\absorbance{}}{}
	 \ifundef\opticdensity{\def\opticdensity{}}{}
	 \ifundef\coulomb{\def\coulomb{}}{}
	 \ifundef\farad{\def\farad{}}{}
	 \ifundef\gram{\def\gram{}}{}
	 \ifundef\gray{\def\gray{}}{}
	 \ifundef\hertz{\def\hertz{}}{}
	 \ifundef\henry{\def\henry{}}{}
	 \ifundef\joule{\def\joule{}}{}
	 \ifundef\katal{\def\katal{}}{}
	 \ifundef\lumen{\def\lumen{}}{}
	 \ifundef\lux{\def\lux{}}{}
	 \ifundef\newton{\def\newton{}}{}
	 \ifundef\ohm{\def\ohm{}}{}
	 \ifundef\pascal{\def\pascal{}}{}
	 \ifundef\radian{\def\radian{}}{}
	 \ifundef\siemens{\def\siemens{}}{}
	 \ifundef\sievert{\def\sievert{}}{}
	 \ifundef\steradian{\def\steradian{}}{}
	 \ifundef\tesla{\def\tesla{}}{}
	 \ifundef\volt{\def\volt{}}{}
	 \ifundef\watt{\def\watt{}}{}
	 \ifundef\weber{\def\weber{}}{}
	 \ifundef\hectare{\def\hectare{}}{}
	 \ifundef\tonne{\def\tonne{}}{}
	 \ifundef\astronomicalunit{\def\astronomicalunit{}}{}
	 \ifundef\atomicmassunit{\def\atomicmassunit{}}{}
	 \ifundef\dalton{\def\dalton{}}{}
	 \ifundef\are{\def\are{}}{}
	 \ifundef\bel{\def\bel{}}{}
	 \ifundef\decibel{\def\decibel{}}{}
	 \ifundef\mmHg{\def\mmHg{}}{}
	 \ifundef\neper{\def\neper{}}{}
	 \ifundef\KWH{\def\KWH{}}{}
	 \ifundef\curie{\def\curie{}}{}
	 \ifundef\gal{\def\gal{}}{}
	 \ifundef\rad{\def\rad{}}{}
	 \ifundef\rem{\def\rem{}}{}
	 \ifundef\roentgen{\def\roentgen{}}{}
	 \ifundef\parsec{\def\parsec{}}{}
	 \ifundef\lightyear{\def\lightyear{}}{}
	 \ifundef\torr{\def\torr{}}{}
	 \ifundef\gon{\def\gon{}}{}
	 \ifundef\gauss{\def\gauss{}}{}
	 \ifundef\bit{\def\bit{}}{}
	 \ifundef\byte{\def\byte{}}{}
}

\RequirePackage{xymtex}
\RequirePackage{physics2}
\let\partial\relax
\DeclareMathSymbol{\partial}{\mathord}{letters}{"40}

\AtEndPreamble{%

\usepackage{hyperref}
\hypersetup{%
	backref,%
	colorlinks=false}
\hypersetup{pdfborder=0 0 0}

\hypersetup{pdfencoding=t2a}

\ifthenelse{\boolean{xetex}}{}{%
	\usepackage{breakurl}
}

\makeatletter
\def\fullref{\@ifstar{\@fullrefstar}{\@fullref}}
\def\@fullrefstar#1{\hyperref[{#1}]{\autoref*{#1} \nameref*{#1}}} 
\def\@fullref#1{\hyperref[{#1}]{\autoref*{#1} \iflanguage{russian}{<<}{``}\nameref*{#1}\iflanguage{russian}{>>}{''}}} 
\makeatother

} 

\usepackage{amsmath}
\usepackage{amssymb}
\usepackage{amsfonts}
\usepackage{amsopn}
\usepackage{slashed}
\usepackage{accents}
\usepackage{amstext}

\DeclareFontFamily{U}{pzc}{\hyphenchar\font45}
\DeclareFontShape{U}{pzc}{m}{it}{%
			<->s*[1.15] pzcmi7t}{}

\DeclareFontFamily{U}{matha}{\hyphenchar\font45}
\DeclareFontShape{U}{matha}{m}{n}{
			<5> <6> <7> <8> <9> <10> gen * matha
			<10.95> matha10 <12> <14.4> <17.28> <20.74> <24.88> matha12
			}{}
\DeclareSymbolFont{matha}{U}{matha}{m}{n}
\DeclareFontSubstitution{U}{matha}{m}{n}

\DeclareFontFamily{U}{mathc}{\hyphenchar\font45}
\DeclareFontShape{U}{mathc}{m}{it}%
{<->s*[1.03] mathc10}{}

\DeclareMathAlphabet{\mathscr}{U}{pzc}{m}{it}
\SetMathAlphabet{\mathscr}{bold}{U}{pzcf}{b}{it}

\DeclareMathAlphabet{\mathcal}{U}{dutchcal}{m}{n}
\SetMathAlphabet{\mathcal}{bold}{U}{dutchcal}{b}{n}

\DeclareMathSymbol{\pm}{3}{matha}{"08}
\DeclareMathSymbol{\mp}{3}{matha}{"09}
\makeatletter
\DeclareMathSymbol{\@cev}{3}{matha}{"D0}
\DeclareMathSymbol{\@vec}{3}{matha}{"D1}
\DeclareMathSymbol{\@vecev}{3}{matha}{"D8}
\def\cev#1{%
	\accentset{\@cev}{\ensuremath{#1}}}
\def\vec#1{%
	\accentset{\@vec}{\ensuremath{#1}}}
\def\vecev#1{%
	\accentset{\@vecev}{\ensuremath{#1}}}
\makeatother

\usepackage{old-arrows}

\usepackage{cancel}

\newcommand\hcancel[2][black]{%
	\ifmmode
		\setbox0=\hbox{$#2$}%
		\mspace{-4mu}%
	\else
		\setbox0=\hbox{#2}%
	\fi
	\rlap{\raisebox{.45\ht0}{\textcolor{#1}{\rule{\wd0}{1pt}}}}#2}

\usepackage{mathtools}
\mathtoolsset{
	mathic=true,
}

\allowdisplaybreaks

\usepackage{xfrac}

\DeclarePairedDelimiter\pqty\lparen\rparen
\DeclarePairedDelimiter\bqty\lbrack\rbrack
\DeclarePairedDelimiter\Bqty\lbrace\rbrace
\DeclarePairedDelimiter\aqty\langle\rangle
\DeclarePairedDelimiter\vqty\lvert\rvert
\DeclarePairedDelimiter\Vqty\lVert\rVert
\DeclarePairedDelimiter\abs\lvert\rvert

\DeclarePairedDelimiter\norm\lVert\rVert

\DeclarePairedDelimiter\scalarproduct\lparen\rparen
\DeclarePairedDelimiter\crossproduct\lbrack\rbrack
\DeclarePairedDelimiter\commutator\lbrack\rbrack
\DeclarePairedDelimiter\anticommutator\lbrace\rbrace

\def\bigO{\mathop{}\!\mathcal{O}}
\def\smallO{%
	\mathchoice
	{
		\mathop{}\!{\scriptstyle\mathcal{O}}
	}
	{
		\mathop{}\!{\scriptstyle\mathcal{O}}
	}
	{
		\mathop{}\!{\scriptscriptstyle\mathcal{O}}
	}
	{
		\mathop{}\!{\scalebox{0.8}{$\scriptscriptstyle\mathcal{O}$}}
	}
}

\makeatletter
\def\bigOp{\@ifnextchar[{\bigOp@opt}{\bigOp@opt[]}}
\def\bigOp@opt[#1]#2{\ensuremath{\bigO\pqty[#1]{#2}}}
\def\bigOb{\@ifnextchar[{\bigOb@opt}{\bigOb@opt[]}}
\def\bigOb@opt[#1]#2{\ensuremath{\bigO\bqty[#1]{#2}}}
\def\bigOB{\@ifnextchar[{\bigOB@opt}{\bigOB@opt[]}}
\def\bigOB@opt[#1]#2{\ensuremath{\bigO\Bqty[#1]{#2}}}
\def\bigOa{\@ifnextchar[{\bigOa@opt}{\bigOa@opt[]}}
\def\bigOa@opt[#1]#2{\ensuremath{\bigO\aqty[#1]{#2}}}
\def\bigOv{\@ifnextchar[{\bigOv@opt}{\bigOv@opt[]}}
\def\bigOv@opt[#1]#2{\ensuremath{\bigO\vqty[#1]{#2}}}
\def\bigOV{\@ifnextchar[{\bigOV@opt}{\bigOV@opt[]}}
\def\bigOV@opt[#1]#2{\ensuremath{\bigO\Vqty[#1]{#2}}}

\def\smallOp{\@ifnextchar[{\smallOp@opt}{\smallOp@opt[]}}
\def\smallOp@opt[#1]#2{\ensuremath{\smallO\pqty[#1]{#2}}}
\def\smallOb{\@ifnextchar[{\smallOb@opt}{\smallOb@opt[]}}
\def\smallOb@opt[#1]#2{\ensuremath{\smallO\bqty[#1]{#2}}}
\def\smallOB{\@ifnextchar[{\smallOB@opt}{\smallOB@opt[]}}
\def\smallOB@opt[#1]#2{\ensuremath{\smallO\Bqty[#1]{#2}}}
\def\smallOa{\@ifnextchar[{\smallOa@opt}{\smallOa@opt[]}}
\def\smallOa@opt[#1]#2{\ensuremath{\smallO\aqty[#1]{#2}}}
\def\smallOv{\@ifnextchar[{\smallOv@opt}{\smallOv@opt[]}}
\def\smallOv@opt[#1]#2{\ensuremath{\smallO\vqty[#1]{#2}}}
\def\smallOV{\@ifnextchar[{\smallOV@opt}{\smallOa@opt[]}}
\def\smallOV@opt[#1]#2{\ensuremath{\smallO\Vqty[#1]{#2}}}

\def\order{\@ifstar{\smallOp}{\bigOp}}
\def\orderp{\@ifstar{\smallOp}{\bigOp}}
\def\orderb{\@ifstar{\smallOb}{\bigOb}}
\def\orderB{\@ifstar{\smallOB}{\bigOB}}
\def\ordera{\@ifstar{\smallOa}{\bigOa}}
\def\orderv{\@ifstar{\smallOv}{\bigOv}}
\def\orderV{\@ifstar{\smallOV}{\bigOV}}
\makeatother

\makeatletter
\def\scalar{\@ifstar{\sc@l@rproduct}{\sc@larproduct}}
\def\cross{\@ifstar{\@cross@product}{\@crossproduct}}
\def\comm{\@ifstar{\@commut@tor}{\commut@tor}}
\def\acomm{\@ifstar{\@nticommut@tor}{\anticommut@tor}}

\def\sc@l@rproduct{\@ifstar{\@sc@l@rproduct}{\sc@l@r@product}}
\def\@cross@product{\@ifstar{\@@cross@product}{\@cross@@product}}

\def\@sc@l@rproduct#1#2{{#1 \cdot #2}}
\def\@@cross@product#1#2{{#1 \times #2}}

\def\sc@larproduct{\@ifnextchar[{\sc@larproduct@opt}{\sc@larproduct@opt[]}}
\def\sc@l@r@product{\@ifnextchar[{\sc@l@rproduct@opt}{\sc@l@rproduct@opt[]}}
\def\@crossproduct{\@ifnextchar[{\@crossproduct@opt}{\@crossproduct@opt[]}}
\def\@cross@@product{\@ifnextchar[{\@cross@product@opt}{\@cross@product@opt[]}}
\def\commut@tor{\@ifnextchar[{\commut@tor@opt}{\commut@tor@opt[]}}
\def\@commut@tor{\@ifnextchar[{\@commut@tor@opt}{\@commut@tor@opt[]}}
\def\anticommut@tor{\@ifnextchar[{\anticommut@tor@opt}{\anticommut@tor@opt[]}}
\def\@nticommut@tor{\@ifnextchar[{\@nticommut@tor@opt}{\@nticommut@tor@opt[]}}

\def\sc@larproduct@opt[#1]#2#3{\scalarproduct[#1]{#2,#3}}
\def\sc@l@rproduct@opt[#1]#2#3{\scalarproduct[#1]{#2 \cdot #3}}
\def\@crossproduct@opt[#1]#2#3{\crossproduct[#1]{#2,#3}}
\def\@cross@product@opt[#1]#2#3{\crossproduct[#1]{#2 \times #3}}
\def\commut@tor@opt[#1]#2#3{\commutator[#1]{#2,#3}}
\def\@commut@tor@opt[#1]#2#3{\commutator[#1]{#2,#3}_{-}}
\def\anticommut@tor@opt[#1]#2#3{\anticommutator[#1]{#2,#3}}
\def\@nticommut@tor@opt[#1]#2#3{\commutator[#1]{#2,#3}_{+}}
\makeatother

\DeclarePairedDelimiter\brange{\lbrack}{\rbrack}

\usepackage[italic=true]{derivative}

\RequirePackage{mleftright}
\derivset{all}[scale-auto=mleftmright]
\derivset{\pdif}[style-var-*=single,style-var=multiple]
\derivset{\fdv}[style-var=multiple,style-var-!=mixed]
\derivset{\mdv}[style-var=multiple,style-var-!=mixed]

\DeclareDifferential{\vdif}{\nabla}[style-var=multiple,style-var-!=mixed]

\DeclareDifferential{\sdif}{\square}[style-var=multiple,style-var-!=mixed]

\DeclareDerivative{\vdv}{\nabla}[style-var=multiple,style-var-!=mixed]

\DeclareDerivative{\sdv}{\square}[style-var=multiple,style-var-!=mixed]

\DeclareDerivative{\edv}{}[fun=false,style-inf=,
style-var=multiple,style-frac=,style-frac-/=,
style-var-/=multiple,style-var-!=single,
style-var-/!=single,delims-eval=(),
delims-eval-/=..,delims-eval-!=..,]

\newlength{\tempheight}

\makeatletter
\newcommand{\vplus}{\mathbin{\mathpalette\vplus@\relax}}
\newcommand{\vplus@}[2]{%
	\raisebox{\depth}{\scalebox{1}[-1]{$\m@th#1\uplus$}}%
}
\newcommand{\bigvplus}{\mathbin{\mathpalette\bigvplus@\relax}}
\newcommand{\bigvplus@}[2]{%
	\raisebox{\depth}{\scalebox{1}[-1]{$\m@th#1\biguplus$}}%
}
\newcommand{\vtimes}{\mathbin{\mathpalette\vtimes@\relax}}
\newcommand{\vtimes@}[2]{%
	\raisebox{\depth}{\scalebox{1}[-1]{$\m@th#1\utimes$}}%
}
\newcommand{\bigvtimes}{\mathbin{\mathpalette\bigvtimes@\relax}}
\newcommand{\bigvtimes@}[2]{%
	\raisebox{\depth}{\scalebox{1}[-1]{$\m@th#1\bigutimes$}}%
}
\newcommand{\vminus}{\mathbin{\mathpalette\vminus@\relax}}
\newcommand{\vminus@}[2]{%
	\raisebox{\depth}{\scalebox{1}[-1]{$\m@th#1\uminus$}}%
}
\newcommand{\bigvminus}{\mathbin{\mathpalette\bigvminus@\relax}}
\newcommand{\bigvminus@}[2]{%
	\raisebox{\depth}{\scalebox{1}[-1]{$\m@th#1\biguminus$}}%
}
\makeatother

\newcommand\utimes{\mathbin{\ooalign{$\cup$\cr%
			\hfil\raise0.42ex\hbox{$\scriptscriptstyle\times$}\hfil\cr}}}
\newcommand\bigutimes{\mathop{\ooalign{$\bigcup$\cr%
			\hfil\raise0.36ex\hbox{$\scriptscriptstyle\boldsymbol{\times}$}\hfil\cr}}}

\newcommand\uminus{\mathbin{\ooalign{$\cup$\cr%
			\hfil\raise0.42ex\hbox{$\scriptscriptstyle-$}\hfil\cr}}}
\newcommand\biguminus{\mathop{\ooalign{$\bigcup$\cr%
			\hfil\raise0.36ex\hbox{$\scriptscriptstyle\boldsymbol{-}$}\hfil\cr}}}

\usepackage{pgfplotstable}
\pgfplotsset{compat=1.8}
\usepackage{booktabs}
\usepackage{array}
\usepackage{stabular}
\usepackage{colortbl}
\usepackage{multirow}
\usepackage{longtable}
\usepackage{makecell}

\pgfplotstableset{/pgf/number format/.append style={
	},
	dec sep={,},
	1000 sep={\,},
	min exponent for 1000 sep=5,
	sci zerofill,
	precision=3,
	read comma as period,
	col sep=semicolon,
}

\usepackage{graphicx}
\usepackage[table,rgb]{xcolor}

\iftexht
	
\else
	\usepackage{float}
\fi

\usepackage{latexcolors}
\definecolor{verysoftblue}{rgb}{0.7,0.7,0.9}
\definecolor{darkteal}{rgb}{0.0,0.2,0.2}
\definecolor{darkyellow}{rgb}{0.5,0.5,0.0}

\graphicspath{{image/}}

\usepackage[scanall]{psfrag}

\usepackage{tikz}
\usetikzlibrary{arrows,shapes,graphs,automata,positioning,calc}
\usetikzlibrary{patterns}
\usetikzlibrary{fadings}

\usepackage{tikz-cd}

\tikzset{terminator/.style={
		rectangle, draw, text centered, rounded corners, minimum height=2em
	}
}

\tikzset{operator/.style={
		rectangle, draw, text centered, minimum height=2em
	}
}

\tikzset{decision/.style={
		diamond, draw, text centered, minimum height=2em
	}
}

\tikzset{data/.style={
		trapezium, draw, text centered, trapezium left angle=60,
		trapezium right angle=120, minimum height=2em
	}
}

\tikzset{link/.style={
		draw, -latex'
	}
}

\tikzset{connector/.style={
		circle, draw, text centered, minimum height=2em
	}
}

\tikzset{function/.style={
		rectangle split, rectangle split horizontal,
		rectangle split parts=3,
		draw, text centered,
		minimum width=5em,
		minimum height=2em,
		outer sep=0
	}
}

\tikzset{subroutine/.style={
		rectangle split, rectangle split horizontal,
		rectangle split parts=3,
		draw, text centered,
		minimum width=5em,
		minimum height=2em,
		outer sep=0
	}
}

\tikzset{procedure/.style={
		rectangle split, rectangle split horizontal,
		rectangle split parts=3,
		draw, text centered,
		minimum width=5em,
		minimum height=2em,
		outer sep=0
	}
}

\tikzset{loop/.style={
		chamfered rectangle,
		chamfered rectangle corners={
			north west, north east
		},
		draw, text centered,
		minimum width=5em,
		minimum height=2em,
		outer sep=0
	}
}

\tikzset{endloop/.style={
		chamfered rectangle,
		chamfered rectangle corners={
			south west, south east
		},
		draw, text centered,
		minimum width=5em,
		minimum height=2em,
		outer sep=0
	}
}

\tikzset{counterloop/.style={
		chamfered rectangle,
		chamfered rectangle xsep=2cm,
		draw, text centered,
		minimum width=5em,
		minimum height=2em,
		outer sep=0
	}
}

\tikzset{preparation/.style={
		rectangle split, rectangle split horizontal,
		rectangle split parts=3,
		draw, text centered,
		minimum width=5em,
		minimum height=2em,
		outer sep=0
	}
}

\RequirePackage{pgfplots}
\pgfplotsset{/pgf/number format/.append style={%
		dec sep={,},
		1000 sep={\,},
		sci zerofill,
		precision=3,
		col sep=semicolon,
	}
	read comma as period,
}
\usepgfplotslibrary{units}
\usepgfplotslibrary{fillbetween}

\makeatletter
\@ifclassloaded{revtex4-1}{\relax}{%
	\@ifclassloaded{beamer}{\relax}{%
		\usepackage{cite}%
	}
}
\makeatother

\usepackage{bibunits}
\bibliographyunit[\title]

\usepackage{listings}
\usepackage{listingsutf8}
\usepackage{listings-ext}
\ifthenelse{\boolean{luatex}\OR\boolean{xetex}}{}{%
		\lstset{inputencoding=utf8/koi8-r}}

\usepackage{siunitx}

\AtBeginDocument%
{%
	\addto\extrasrussian%
	{
		\input{siunitx_cyr}%
	}
	\addto\extrasenglish%
	{
		\input{siunitx_lat}%
	}
}

\RequirePackage{xkeyval}
\usepackage{physics2}
\usephysicsmodule[tightbraces=true]{ab}

\usepackage{braket}

\makeatletter
\def\expval{\@ifstar{\expv@l}{\@expval}}

\def\@expval#1{\@ifnextchar\bgroup{\@@expval{#1}}{\Braket{#1}}}
\def\@@expval#1#2{\Braket{{#2}|#1|{#2}}}

\def\expv@l#1{\@ifnextchar\bgroup{\@expv@l{#1}}{\braket{#1}}}
\def\@expv@l#1#2{\braket{{#2}|#1|{#2}}}

\def\ev{\@ifstar{\expect@tionv@lue}{\expect@tionvalue}}
\let\expect@tionvalue=\Braket
\let\expect@tionv@lue=\braket

\def\matrixel{\@ifstar{\m@trixel}{\@matrixel}}
\def\@matrixel#1#2#3{\Braket{{#1}|{#2}|{#3}}}
\def\m@trixel#1#2#3{\braket{{#1}|{#2}|{#3}}}

\def\mel{\@ifstar{\@m@trixelement}{\@matrixelement}}
\let\@matrixelement=\Braket
\let\@m@trixelement=\braket
\makeatother

\usepackage{xymtex}

\usepackage{prooftrees}

\usepackage{xstring}
\usepackage{listofitems}

\usepackage{readarray}

\usepackage[heightadjust]{marginnote}

\AtBeginDocument{%
	\newdimen\marginwidth
	\setlength{\marginwidth}{\paperwidth-\linewidth}
	\setlength{\marginwidth}{0.8\marginwidth/2}
	\xdef\marginparwidth{\marginwidth}
	\xdef\marginparsep{5pt}%
}

\RequirePackage{ulem}
\RequirePackage{soul}
\RequirePackage{ifthen}
\RequirePackage{xcolor}
\RequirePackage[loadshadowlibrary,shadow]{todonotes}
\RequirePackage{readarray}
\RequirePackage{tcolorbox}
\RequirePackage{shellesc}
\RequirePackage{latexcolors}

\readarraysepchar{;}

\makeatletter
\def\@default@fgcolor{black}
\def\@default@bgcolor{white}
\def\@fixme@color{red}
\def\@alert@color{scarlet}
\def\@highlight@color{yellow}
\def\@edit@color{darkyellow}
\def\@add@color{blue}
\def\@remove@color{crimson}
\def\@comment@color{midnightblue}
\def\@comment@fgcolor{darkgreen}
\def\@comment@bgcolor{aliceblue}
\def\@comment@fcolor{firebrick}
\def\@commented@fgcolor{darkgreen}
\def\@commented@bgcolor{aliceblue}
\def\@commented@fcolor{firebrick}

\def\coloreditor@page@num{\thepage}%

\newif\ifistoreview
\newif\ifistosupressedits
\newif\ifistoshowdeletions

\istoreviewtrue
\istosupresseditsfalse
\istoshowdeletionstrue

\newif\ifis@torotatecolors
\is@torotatecolorsfalse

\newcommand{\coloreditor@ProcessOptions}{%
	\ifistosupressedits
		\def\coloreditor@SuppressEdits{yes}
	\else
		\def\coloreditor@SuppressEdits{no}
	\fi
	\ifistoshowdeletions
		\def\coloreditor@ShowDeletions{yes}
	\else
		\def\coloreditor@ShowDeletions{no}
	\fi
}

\def\coloreditor@clr@cnt{5}

\def\coloreditor@removedmessage{здесь удалено}%

\AtBeginDocument{%
	\addto\extrasrussian{%
		\def\coloreditor@removedmessage{здесь удалено}%
	}%
	\addto\extrasenglish{%
		\def\coloreditor@removedmessage{removed here}%
	}%
}

\def\coloreditor@chiefeditorName{Главред}%

\AtBeginDocument{%
	\addto\extrasrussian{%
		\def\coloreditor@chiefeditorName{Главред}%
	}%
	\addto\extrasenglish{%
		\def\coloreditor@chiefeditorName{ChiefEd}%
	}%
}

\newcounter{coloreditor@clr@srs@num}%
\newcounter{coloreditor@auth@id@num}%
\newdimen\clred@gen@cmt@i@x
\newdimen\clred@gen@cmt@i@y
\newdimen\clred@gen@cmt@s@x
\newdimen\clred@gen@cmt@s@y

\newdimen\cmt@str@len

\newdimen\epsilon@y
\global\newdimen\cur@pos@x
\global\newdimen\cur@pos@y

\def\hl@tcb@bottom{1pt}
\def\hl@tcb@top{2.5pt}
\def\hl@tcb@right{1pt}
\def\hl@tcb@left{1pt}
\def\hl@tcb@boxsep{0pt}
\def\hl@tcb@boxrule{1pt}
\def\hl@tcb@defaultcolframe{yellow}

\newtcbox{\hlbox}[2][\hl@tcb@defaultcolframe]{on line,arc=\hl@tcb@arc,colback=#2,colframe=#1,boxrule=\hl@tcb@boxrule,boxsep=\hl@tcb@boxsep,left=\hl@tcb@left,right=\hl@tcb@right,top=\hl@tcb@top,bottom=\hl@tcb@bottom,sharp corners=\hl@tcb@sharpcorners}%

\tikzstyle{notestyleraw} = [%
	draw=\@todonotes@currentbordercolor,%
	fill=\@todonotes@currentbackgroundcolor,%
	text=\@todonotes@currenttextcolor,%
	line width=0.8pt,%
	text width = \@todonotes@textwidth - 1.6 ex - 1pt,%
	inner sep = 0.8 ex,%
	rounded corners=4pt]%

\tikzstyle{connectstyle} = [%
	thick,%
	draw=\@todonotes@currentlinecolor]%

\renewcommand{\@todonotes@drawMarginNoteWithLine}{%
	\ifvmode
		\vspace*{-\parskip}		
		\vskip-\baselineskip	
		\noindent
	\fi
	\begin{tikzpicture}[remember picture, overlay, baseline={-(\hl@tcb@bottom+\hl@tcb@boxrule+\hl@tcb@boxsep-0.06ex)}]
		\node [coordinate] (inText) {};%
	\end{tikzpicture}%
	\marginpar[{
		\@todonotes@drawMarginNote%
		\@todonotes@drawLineToLeftMargin%
	}]{
		\@todonotes@drawMarginNote%
		\@todonotes@drawLineToRightMargin%
	}%
}%

\newcommand{\coloreditor@addeditorremovecounter}[1]{%
	\newcounter{#1@cnt@remove}[page]%
	\setcounter{#1@cnt@remove}{0}%
}

\newcommand{\coloreditor@addeditormargincommentcoordinates}[1]{%
	\global\newcounter{#1@cmt@first@was}[page]%
	\setcounter{#1@cmt@first@was}{0}%
	\expandafter\newdimen\csname clred#1@cmt@i@x@prev\endcsname
	\expandafter\newdimen\csname clred#1@cmt@i@y@prev\endcsname
	\expandafter\newdimen\csname clred#1@cmt@s@x@prev\endcsname
	\expandafter\newdimen\csname clred#1@cmt@s@y@prev\endcsname
	\expandafter\newdimen\csname clred#1@cmt@i@x@cur\endcsname
	\expandafter\newdimen\csname clred#1@cmt@i@y@cur\endcsname
	\expandafter\newdimen\csname clred#1@cmt@s@x@cur\endcsname
	\expandafter\newdimen\csname clred#1@cmt@s@y@cur\endcsname
	\expandafter\newdimen\csname clred#1@cmt@distance@y\endcsname
}

\def\clred@obj@names{comment,commented}%
\def\clred@obj@type{fg,bg}%

\newcommand{\coloreditor@addeditorcolors}[7]{%
	\expandafter\long\expandafter\edef\csname #1@edit@color\endcsname{#2}%
	\expandafter\long\expandafter\edef\csname #1@add@color\endcsname{#3}%
	\expandafter\long\expandafter\edef\csname #1@remove@color\endcsname{#4}%
	\expandafter\long\expandafter\edef\csname #1@comment@color\endcsname{#5}%
	\coloreditor@addeditorcolorseries{#1}{#6}{#7}%
	\newcounter{clred#1@clr@num}%
	\def\do@def##1##2{%
		\ifthenelse{\equal{##2}{fg}}{\edef\crled@arg@iv{#6}}{\edef\crled@arg@iv{#7}}%
		\expandafter\long\expandafter\edef\csname #1@##1@##2color\endcsname{\crled@arg@iv}%
		\coloreditor@addeditorcolorrotations{#1}{##1}{##2}{\crled@arg@iv}%
	}%
	\@for\next@i:=\clred@obj@names\do{%
		\@for\next@ii:=\clred@obj@type\do{%
			\expandafter\do@def\expandafter{\next@i}{\next@ii}%
		}%
	}%
}

\newcommand{\coloreditor@addeditorauth}[1]{%
	\expandafter\long\expandafter\def\csname clred#1@auth@id\endcsname{#1}%
	\expandafter\long\expandafter\def\csname clred#1@auth@num\endcsname{\thecoloreditor@auth@id@num}%
	\stepcounter{coloreditor@auth@id@num}%
}

\newcommand{\coloreditor@addeditorcolorseries}[3]{%
	\definecolorseries{#1@clr@fg@modified}{cmy}{grad}{#2}[hsb]{757.07,137,426}%
	\resetcolorseries{#1@clr@fg@modified}%
	\definecolorseries{#1@clr@bg@modified}{cmy}{grad}{#3}[hsb]{0.01,8.01,0.0426}%
	\resetcolorseries{#1@clr@bg@modified}%
}

\newcommand{\coloreditor@addeditorcolorrotations}[4]{%
	\expandafter\long\expandafter\edef\csname #1@#2@#3color@i\endcsname{#4}%
	\setcounter{clred#1@clr@num}{1}%
	\loop\unless\if\csname theclred#1@clr@num\endcsname\coloreditor@clr@cnt
		\stepcounter{clred#1@clr@num}%
		\setcounter{coloreditor@clr@srs@num}{\coloreditor@clr@cnt-\csname theclred#1@clr@num\endcsname+2}
		\expandafter\long\expandafter\edef\csname #1@#2@#3color@\romannumeral\thecoloreditor@clr@srs@num\endcsname{#1@clr@#3@modified!![\csname theclred#1@clr@num\endcsname]}%
	\repeat
	\setcounter{clred#1@clr@num}{1}%
}

	\def\coordinate@marks{i,s}%
	\def\coordinate@projections{x,y}%

\newcommand{\coloreditor@rotatecolors@auto}[1]{%
	\is@torotatecolorsfalse
	\edef\chknum@i{\csname the#1@cmt@first@was\endcsname}%
	\edef\chknum@ii{0}%
	\ifnum\chknum@i=\chknum@ii
		\setcounter{#1@cmt@first@was}{1}%
	\else
		\def\do@def##1##2{%
			\coloreditor@coordinates@save@from@to{clred#1@cmt@##1@##2@cur}{clred#1@cmt@##1@##2@prev}%
		}%
		\@for\next@i:=\coordinate@marks\do{%
			\@for\next@ii:=\coordinate@projections\do{%
				\expandafter\do@def\expandafter{\next@i}{\next@ii}%
			}%
		}%
	\fi
	\edef\chknum@ii{1}%
	\def\do@def##1##2{%
		\coloreditor@coordinates@save@from@to{clred@gen@cmt@##1@##2}{clred#1@cmt@##1@##2@cur}%
	}%
	\@for\next@i:=\coordinate@marks\do{%
		\@for\next@ii:=\coordinate@projections\do{%
			\expandafter\do@def\expandafter{\next@i}{\next@ii}%
		}%
	}%
	\ifnum\chknum@i=\chknum@ii
		\def\chkdim@i{\csname clred#1@cmt@i@y@cur\endcsname}%
		\def\chkdim@ii{\csname clred#1@cmt@s@y@prev\endcsname}%
		\def\chkdim@iii{\csname clred#1@cmt@distance@y\endcsname}%
		\setlength{\chkdim@iii}{\dimexpr\chkdim@ii-\chkdim@i\relax}%
		\ifdim\the\chkdim@iii<0pt
			\setlength{\chkdim@iii}{\dimexpr-\chkdim@iii\relax}%
		\fi
		\ifdim\chkdim@iii<\epsilon@y
			\is@torotatecolorstrue%
		\else
			\is@torotatecolorsfalse%
		\fi
	\fi
	\ifis@torotatecolors
		\coloreditor@rotatecolors{#1}%
	\else
		\edef\chkclrnum@i{\csname theclred#1@clr@num\endcsname}%
		\edef\chkclrnum@ii{1}%
		\ifnum\chkclrnum@i>\chkclrnum@ii
			\coloreditor@setcolors@default{#1}%
		\fi
	\fi
}

\newcommand{\coloreditor@coordinates@save@from@to}[2]{%
	\setlength{\csname #2\endcsname}{\csname #1\endcsname}%
}

\newcommand{\coloreditor@rotatecolors}[1]{%
	\coloreditor@rotatecolors@chkcnt{#1}%
	\coloreditor@setcolors{#1}{comment}%
	\coloreditor@setcolors{#1}{commented}%
}

\newcommand{\coloreditor@rotatecolors@chkcnt}[1]{%
	\edef\chkclrnum@i{\csname theclred#1@clr@num\endcsname}%
	\edef\chkclrnum@ii{\coloreditor@clr@cnt}%
	\ifnum\chkclrnum@i<\chkclrnum@ii
		\stepcounter{clred#1@clr@num}%
	\else
		\setcounter{clred#1@clr@num}{1}%
	\fi
}

\newcommand{\coloreditor@setcolors}[2]{%
	\expandafter\long\expandafter\edef\csname #1@#2@fgcolor\endcsname{\csname #1@#2@fgcolor@\romannumeral\csname theclred#1@clr@num\endcsname\endcsname}%
	\expandafter\long\expandafter\edef\csname #1@#2@bgcolor\endcsname{\csname #1@#2@bgcolor@\romannumeral\csname theclred#1@clr@num\endcsname\endcsname}%
}

\newcommand{\coloreditor@setcolors@default}[1]{%
	\setcounter{clred#1@clr@num}{1}%
	\coloreditor@setcolors{#1}{comment}%
	\coloreditor@setcolors{#1}{commented}%
}

\newcommand{\coloreditor@addeditoredit@aux}[2]{%
	\coloreditor@addeditoredit{#1}{#2}{\csname #1@edit@color\endcsname}%
}

\newcommand{\coloreditor@addeditoredit}[3]{%
	\ifthenelse{\equal{\coloreditor@SuppressEdits}{yes} \or \equal{#2}{suppress}}{
		\expandafter\long\expandafter\def\csname #1edit\endcsname ##1{##1}%
	}{
		\expandafter\long\expandafter\def\csname #1edit\endcsname ##1{%
			\ifistoreview
				\coloring{#3}{##1}%
			\else ##1\fi}
	}}

\newcommand{\coloreditor@addeditoradd@aux}[2]{%
	\coloreditor@addeditoradd{#1}{#2}{\csname #1@add@color\endcsname}%
}

\newcommand{\coloreditor@addeditoradd}[3]{%
	\ifthenelse{\equal{\coloreditor@SuppressEdits}{yes} \or \equal{#2}{suppress}}{
		\expandafter\long\expandafter\def\csname #1add\endcsname ##1{##1}%
	}{
		\expandafter\long\expandafter\def\csname #1add\endcsname ##1{%
			\ifistoreview
				\coloring{#3}{##1}%
			\else ##1\fi}
	}}

\newsavebox{\toremove@box}

\newcommand{\coloreditor@addeditorremove@aux}[2]{%
	\coloreditor@addeditorremove{#1}{#2}{\csname #1@remove@color\endcsname}%
}

\newcommand{\coloreditor@addeditorremove}[3]{%
	\ifthenelse{\equal{\coloreditor@SuppressEdits}{yes} \or \equal{#2}{suppress}}{
		\expandafter\long\expandafter\def\csname #1remove\endcsname{%
			\@ifnextchar[%
				{\csname #1remove@aux\endcsname}%
				{\csname #1remove@aux\endcsname []}}%
		\expandafter\long\expandafter\def\csname #1remove@aux\endcsname [##1]##2{}%
	}{
		\expandafter\long\expandafter\def\csname #1remove\endcsname{%
			\@ifnextchar[%
				{\csname #1remove@aux\endcsname}%
				{\csname #1remove@aux\endcsname []}}%
		\ifthenelse{\equal{\coloreditor@ShowDeletions}{yes}}{
			\expandafter\long\expandafter\def\csname #1remove@aux\endcsname [##1]##2{%
				\ifistoreview
					\ifmmode
					{%
						\savebox{\toremove@box}{\ensuremath{##2}}%
						\ensuremath{\mathcolor{#3}{%
								\ooalign{%
									\hidewidth ##2\hidewidth\cr\rule[0.5ex]{\wd\toremove@box}{0.4pt}%
								}%
						}}%
					}%
					\else{\color{#3}\sout{##2}}\fi
				\else\unskip\fi}%
		}{
			\expandafter\long\expandafter\def\csname #1remove@aux\endcsname [##1]##2{%
				\ifistoreview
					\stepcounter{#1@cnt@remove}%
					{\color{#3}<\csname the#1@cnt@remove\endcsname>\marginpar{%
						\scriptsize\color{#3}#2 <\csname the#1@cnt@remove\endcsname>%
						\coloreditor@removedmessage{%
							\ifthenelse{\equal{##1}{}}{}{: ##1}}%
						}}%
				\fi}%
		}%
	}}

\newcommand{\coloreditor@addeditorreplace}[2]{%
	\ifthenelse{\equal{\coloreditor@SuppressEdits}{yes} \or \equal{#2}{suppress}}{
		\expandafter\long\expandafter\def\csname #1replace\endcsname ##1##2{##2}
	}{
		\ifthenelse{\equal{\coloreditor@ShowDeletions}{yes}}{
			\expandafter\long\expandafter\def\csname #1replace\endcsname ##1##2{%
				\ifistoreview
					\csname #1remove\endcsname{##1}~\csname #1add\endcsname{##2}%
				\else ##2\fi}%
		}{
			\expandafter\long\expandafter\def\csname #1replace\endcsname ##1##2{%
				\ifistoreview
					\csname #1add\endcsname{##2}%
				\else ##2\fi}%
		}%
	}}

\newcommand{\coloreditor@addeditorcmt@aux}[2]{%
	\coloreditor@addeditorcmt{#1}{#2}{\csname #1@comment@color\endcsname}{\csname #1@comment@fgcolor\endcsname}{\csname #1@comment@bgcolor\endcsname}%
}

\newcommand{\coloreditor@addeditorcmt}[5]{%
	\ifthenelse{\equal{\coloreditor@SuppressEdits}{yes} \or \equal{#2}{suppress}}{
		\expandafter\long\expandafter\def\csname #1cmt\endcsname{%
			\@ifstar%
				{\csname #1@cmt@aux\endcsname}%
				{\csname #1@cmt\endcsname}}%
		\expandafter\long\expandafter\def\csname #1@cmt\endcsname ##1{}%
		\expandafter\long\expandafter\def\csname #1@cmt@aux\endcsname{%
			\@ifstar%
				{\csname #1@cmt@doublestarred\endcsname}%
				{\csname #1@cmt@starred\endcsname}}%
		\expandafter\long\expandafter\def\csname #1@cmt@starred\endcsname{%
			\@ifnextchar[%
				{\csname #1@cmt@starred@aux\endcsname}%
				{\csname #1@cmt@starred@aux\endcsname[noauthor]}}%
		\expandafter\long\expandafter\def\csname #1@cmt@doublestarred\endcsname{%
			\@ifnextchar[%
				{\csname #1@cmt@doublestarred@aux\endcsname}%
				{\csname #1@cmt@doublestarred@aux\endcsname[noauthor]}}%
		\expandafter\long\expandafter\def\csname #1@cmt@starred@aux\endcsname [##1]##2{}%
		\expandafter\long\expandafter\def\csname #1@cmt@doublestarred@aux\endcsname [##1]##2{}%
	}{
		\expandafter\long\expandafter\def\csname #1cmt\endcsname{%
			\@ifstar%
				{\csname #1@cmt@aux\endcsname}%
				{\csname #1@cmt\endcsname}}%
		\expandafter\long\expandafter\def\csname #1@cmt\endcsname ##1{\cmtx[#3]{[#2: ##1]}}%
		\expandafter\long\expandafter\def\csname #1@cmt@aux\endcsname{%
			\@ifstar%
				{\csname #1@cmt@doublestarred\endcsname}%
				{\csname #1@cmt@starred\endcsname}}%
		\expandafter\long\expandafter\def\csname #1@cmt@starred\endcsname{%
			\@ifnextchar[%
				{\csname #1@cmt@starred@aux\endcsname}%
				{\csname #1@cmt@starred@aux\endcsname[noauthor]}}%
		\expandafter\long\expandafter\def\csname #1@cmt@doublestarred\endcsname{%
			\@ifnextchar[%
				{\csname #1@cmt@doublestarred@aux\endcsname}%
				{\csname #1@cmt@doublestarred@aux\endcsname[noauthor]}}%
		\expandafter\long\expandafter\def\csname #1@cmt@starred@aux\endcsname [##1]##2{{\cmtx*[#4]{#2: ##2}[#5][##1]}}%
		\expandafter\long\expandafter\def\csname #1@cmt@doublestarred@aux\endcsname [##1]##2{%
			\ifistoreview
				\coloreditor@margincomment@test{#2: ##2}%
				\coloreditor@rotatecolors@auto{#1}%
			\fi
			{\cmtx*[#4]{#2: ##2}[#5][##1]}}%
	}}

\def\cmtx{\@ifstar{\cmtx@starred}{\@cmtx}}

\def\@cmtx{%
	\@ifnextchar[%
		{\@cmtx@i}%
		{\@cmtx@i[\@comment@color]}%
}

\def\@cmtx@i[#1]#2{\ifistoreview{\color{#1}#2}\fi}

\def\cmtx@starred{%
	\@ifnextchar[%
		{\cmtx@starred@i}%
		{\cmtx@starred@i[\@comment@fgcolor]}%
}
\def\cmtx@starred@i[#1]{%
	\@ifnextchar\bgroup%
		{\cmtx@starred@ii{#1}}%
		{\cmtx@starred@ii{#1}{}}%
}
\def\cmtx@starred@ii#1#2{%
	\@ifnextchar[%
		{\cmtx@starred@iii{#1}{#2}}%
		{\cmtx@starred@iii{#1}{#2}[\@comment@bgcolor]}%
}
\def\cmtx@starred@iii#1#2[#3]{%
	\@ifnextchar[%
		{\cmtx@starred@iv{#1}{#2}{#3}}%
		{\cmtx@starred@iv{#1}{#2}{#3}[noauthor]}%
}
\def\cmtx@starred@iv#1#2#3[#4]{%
	\ifistoreview
		\todo[textcolor=#1,backgroundcolor=#3,bordercolor=#3,linecolor=#3,list,tickmarkheight=1.3ex,line,noinline,#4]{#2}%
	\fi
}

\newcommand{\coloreditor@margincomment@test}[1]{%
	\StrLen{#1}[\cmt@str@char@cnt]%
	\setlength{\cmt@str@len}{\dimexpr\cmt@str@char@cnt em\relax}%
	\marginpar{\storepos}\store@pos%
	\savepos{\clred@gen@cmt@i@x}{\clred@gen@cmt@i@y}%
	\setlength{\clred@gen@cmt@s@x}{\dimexpr\clred@gen@cmt@i@x+\marginwidth\relax}%
	\setlength{\clred@gen@cmt@s@y}{\dimexpr1.5ex\relax}%
	\setlength{\clred@gen@cmt@s@y}{\dimexpr\clred@gen@cmt@i@y-(\clred@gen@cmt@s@y*\cmt@str@len/\marginwidth)\relax}%
}

\newcommand{\coloreditor@addeditorcommented@aux}[2]{%
	\coloreditor@addeditorcommented{#1}{#2}{\csname #1@commented@bgcolor\endcsname}%
	{\csname #1@commented@fgcolor\endcsname}%
}

\newcommand{\coloreditor@addeditorcommented}[4]{%
		\ifthenelse{\equal{\coloreditor@SuppressEdits}{yes} \or \equal{#2}{suppress}}{
		\expandafter\long\expandafter\def\csname #1hls\endcsname{%
			\@ifstar%
				{\csname #1@hls\endcsname}%
				{\csname #1@hls\endcsname}}%
		\expandafter\long\expandafter\def\csname #1@hls\endcsname ##1{##1}%
	}{
		\expandafter\long\expandafter\def\csname #1hls\endcsname{%
			\@ifstar%
				{\csname #1@hls@aux\endcsname}%
				{\csname #1@hls\endcsname}}%
		\expandafter\long\expandafter\def\csname #1@hls@aux\endcsname{%
			\@ifstar%
				{\csname #1@hls@doublestarred\endcsname}%
				{\csname #1@hls@starred\endcsname}}%
		\expandafter\long\expandafter\def\csname #1@hls@starred\endcsname{%
			\ifistoreview
				\coloreditor@setcolors@default{#1}%
			\fi
			\csname #1@hls\endcsname}%
		\expandafter\long\expandafter\def\csname #1@hls@doublestarred\endcsname{%
			\ifistoreview
				\coloreditor@rotatecolors{#1}%
			\fi
			\csname #1@hls\endcsname}%
		\expandafter\long\expandafter\def\csname #1@hls\endcsname ##1{%
			\ifistoreview
				{\highlight[#3][#4]{##1}}%
			\else ##1\fi}%
	}}

\newcommand{\coloreditor@addeditorcomment@aux}[2]{%
	\coloreditor@addeditorcomment{#1}{#2}%
}

\newcommand{\coloreditor@addeditorcomment}[2]{%
	\ifthenelse{\equal{\coloreditor@SuppressEdits}{yes} \or \equal{#2}{suppress}}{
		\expandafter\long\expandafter\def\csname #1comment\endcsname{%
			\@ifnextchar[%
				{\csname #1comment@aux\endcsname}%
				{\csname #1comment@aux\endcsname[noauthor]}}%
		\expandafter\long\expandafter\def\csname #1comment@aux\endcsname [##1]##2##3{##2}%
	}{
		\expandafter\long\expandafter\def\csname #1comment\endcsname{%
			\@ifnextchar[%
				{\csname #1comment@aux\endcsname}%
				{\csname #1comment@aux\endcsname[noauthor]}}%
		\expandafter\long\expandafter\def\csname #1comment@aux\endcsname [##1]##2##3{%
			\ifodd\coloreditor@page@num
				\ifistoreview
					\coloreditor@margincomment@test{#2: ##3}%
					\coloreditor@rotatecolors@auto{#1}%
				\fi
				\csname #1hls\endcsname {##2}\csname #1cmt\endcsname*[##1]{##3}%
			\else
				\if@afterindent\else
					\smash{\llap{}}\hspace*{-0.5em}%
				\fi
				\csname #1cmt\endcsname**[##1]{##3}\csname #1hls\endcsname {##2}%
			\fi}%
	}}

\newcommand{\addeditor}[8][]{%
	\coloreditor@ProcessOptions%
	\coloreditor@addeditorcolors{#2}{#3}{#4}{#5}{#6}{#7}{#8}%
	\coloreditor@addeditorauth{#2}%
	\coloreditor@addeditorremovecounter{#2}%
	\coloreditor@addeditormargincommentcoordinates{#2}%
	\coloreditor@addeditoredit@aux{#2}{#1}%
	\coloreditor@addeditoradd@aux{#2}{#1}%
	\coloreditor@addeditorreplace{#2}{#1}%
	\ifthenelse{\equal{#1}{}}{
		\coloreditor@addeditorcommented@aux{#2}{#2}%
		\coloreditor@addeditorcmt@aux{#2}{#2}%
		\coloreditor@addeditorcomment@aux{#2}{#2}%
		\coloreditor@addeditorremove@aux{#2}{#2}%
	}{
		\coloreditor@addeditorcmt@aux{#2}{#1}%
		\coloreditor@addeditorcommented@aux{#2}{#1}%
		\coloreditor@addeditorcomment@aux{#2}{#1}%
		\coloreditor@addeditorremove@aux{#2}{#1}%
	}%
}

\AtBeginDocument{%
	\addeditor[\coloreditor@chiefeditorName]{}{darkyellow}{blue}{crimson}{midnightblue}{darkgreen}%
	{aliceblue}%
}

\def\storepos{%
	\begin{tikzpicture}[remember picture]
		\coordinate (x) at (0, 0);
		\path[overlay] let \p1=($(x)-(current page.south west)$)
										in \pgfextra{%
													\global\edef\cur@pos@x@str{\x1}%
													\global\edef\cur@pos@y@str{\y1}};
	\end{tikzpicture}\store@pos}

\def\store@pos{%
	\ifdefined\cur@pos@x@str%
		\setlength{\cur@pos@x}{\cur@pos@x@str}%
	\fi
	\ifdefined\cur@pos@y@str%
		\setlength{\cur@pos@y}{\cur@pos@y@str}%
	\fi}

\def\savepos#1#2{%
	\ifdefined\cur@pos@x@str%
		\setlength{#1}{\cur@pos@x@str}%
	\fi
	\ifdefined\cur@pos@y@str%
		\setlength{#2}{\cur@pos@y@str}%
	\fi}

\def\getpos#1{%
	\ifstrequal{#1}{x}{\cur@pos@x}%
	{\ifstrequal{#1}{y}{\cur@pos@y}{}}}

\def\getposstr#1{%
	\ifstrequal{#1}{x}{%
		\ifdefined\cur@pos@x@str%
			\cur@pos@x@str%
		\fi%
	}%
	{\ifstrequal{#1}{y}{%
		\ifdefined\cur@pos@y@str%
			\cur@pos@y@str%
		\fi}{}}}

\def\coloring#1#2{%
	\ifmmode
		{\ensuremath{\mathcolor{#1}{#2}}}%
	\else
		{\color{#1}{#2}}%
	\fi
}

\def\fcoloring#1#2#3#4{%
	\ifmmode
		\mathchoice%
			{
				{\hlbox[#1]{#2}{\ensuremath{\displaystyle\mathcolor{#3}{#4}}}}%
			}%
			{
				{\hlbox[#1]{#2}{\ensuremath{\textstyle\mathcolor{#3}{#4}}}}%
			}%
			{
				{\hlbox[#1]{#2}{\ensuremath{\scriptstyle\mathcolor{#3}{#4}}}}%
			}%
			{
				{\hlbox[#1]{#2}{\ensuremath{\scriptscriptstyle\mathcolor{#3}{#4}}}}%
			}%
	\else
		{\hlbox[#1]{#2}{{\color{#3}#4}}}%
	\fi
}

\def\bcoloring#1#2#3{%
	\ifmmode
		\edef\fboxsep@cur{\fboxsep}%
		\setlength{\fboxsep}{2pt}%
		\mathchoice%
			{
				{\colorbox{#1}{\ensuremath{\displaystyle\mathcolor{#2}{#3}}}}%
			}%
			{
				{\colorbox{#1}{\ensuremath{\textstyle\mathcolor{#2}{#3}}}}%
			}%
			{
				{\colorbox{#1}{\ensuremath{\scriptstyle\mathcolor{#2}{#3}}}}%
			}%
			{
				{\colorbox{#1}{\ensuremath{\scriptscriptstyle\mathcolor{#2}{#3}}}}%
			}%
		\setlength{\fboxsep}{\fboxsep@cur}%
	\else
	{%
		\edef\SOUL@hlcolor@cur{\SOUL@hlcolor}%
		\sethlcolor{#1}{\color{#2}\hl{#3}}%
		\sethlcolor{\SOUL@hlcolor@cur}%
	}%
	\fi
}

\DeclareRobustCommand{\fixme}[1]{\coloring{\@fixme@color}{#1}}

\@ifclassloaded{beamer}%
	{%
		\let\beamer@lert=\alert%
		\def\alert{\@ifstar{\@lert}{\beamer@lert}}%
		\DeclareRobustCommand{\@lert}[1]{\coloring{\@alert@color}{#1}}%
	}%
	{\DeclareRobustCommand{\alert}[1]{\coloring{\@alert@color}{#1}}}%

\ifthenelse{\boolean{pdftex}}%
{%
	\def\highlight{%
		\@ifnextchar[%
			{\highlight@i}%
			{\highlight@i[\@highlight@color]}%
	}
	\def\highlight@i[#1]{%
		\@ifnextchar[%
			{\highlight@ii{#1}}%
			{\highlight@ii{#1}[\@default@fgcolor]}%
	}
	\def\highlight@ii#1[#2]#3{%
		\bcoloring{#1}{#2}{#3}%
	}
}%
{%
	\NewDocumentCommand{\highlight}{ O{\@highlight@color} O{\@default@fgcolor} m }{%
		\bcoloring{#1}{#2}{#3}%
	}
}

\newdimen\str@len

\DeclareRobustCommand{\justifyme}[2][4.5]{%
	\StrLen{#2}[\str@char@cnt]%
	\StrCount{#2}{ }[\str@space@char@cnt]%
	\setlength{\str@len}{\dimexpr\cmt@str@char@cnt em\relax}%
	\setlength{\str@len}{\dimexpr\str@len*\str@char@cnt}%
	{\raggedright\spaceskip #1\fontdimen2 \font #2}%
}
\makeatother

\usepackage{bm}
\usepackage{indentfirst}
\usepackage{epigraph}
\usepackage{textcomp}
\usepackage{cite}
\usepackage{layout}
\usepackage{varwidth}

\usepackage{longtable}
\usepackage{lscape}
\usepackage{epic}
\usepackage[a4,center]{crop}

\usepackage{calc}
\usepackage{soul}
\usepackage{ulem}
\usepackage{ragged2e}

\usepackage{rotating}

\usepackage[scanall]{psfrag}

\usepackage[nodayofweek]{datetime}
\usepackage[russian,calc]{datetime2}

\usepackage{texnames}
\usepackage[defblank,cfg]{paralist}

\usepackage[hyphens]{url}

\usepackage[super,negative]{nth}

\renewcommand{\abstractname}{}
\renewcommand{\abstractname}{}
\makeatletter
\newcommand{\const}{\@ifstar{\c@nst}{\ensuremath{\mathrm{const}}}}
\newcommand{\c@nst}{\ensuremath{\mathrm{\cyrp\cyro\cyrs\cyrt.}}}
\newcommand{\diag}{\@ifstar{\@di@g}{\di@g}}
\DeclareMathOperator*{\@di@g}{\ensuremath{\textrm{\cyrd\cyri\cyra\cyrg}}}
\DeclareMathOperator*{\di@g}{\mathrm{diag}}
\newcommand{\Diag}{\@ifstar{\@Di@g}{\Di@g}}
\DeclareMathOperator*{\@Di@g}{\ensuremath{\textrm{\CYRD\cyri\cyra\cyrg}}}
\DeclareMathOperator*{\Di@g}{\mathrm{Diag}}
\AtBeginDocument{%
	\addto\extrasrussian%
	{%
		\renewcommand{\c@nst}{\ensuremath{\textrm{пост.}}}%
		\renewcommand{\diag}{\@ifstar{\@di@g}{\di@g}}%
		\renewcommand{\Diag}{\@ifstar{\@Di@g}{\Di@g}}%
	}
	\addto\extrasenglish%
	{%
		\renewcommand{\c@nst}{\ensuremath{\mathrm{const}}}%
		\renewcommand{\diag}{\@ifstar{\di@g}{\di@g}}%
		\renewcommand{\Diag}{\@ifstar{\Di@g}{\Di@g}}%
	}
}
\makeatother

\DeclareMathOperator{\PV}{\mathrm{P.\kern -0.1em V.}}

\DeclareMathOperator{\sgn}{\mathrm{sgn}}

\DeclareMathOperator{\gradient}{\nabla}
\DeclareMathOperator{\divergence}{\nabla\,\cdot}
\DeclareMathOperator{\rotor}{\nabla\,\times}
\DeclareMathOperator{\conditionnumber}{\kappa}
\DeclareMathOperator{\laplaceoperator}{\Delta}

\let\div\relax

\makeatletter
\def\grad{\@ifstar{\@gradient}{\gradient}}
\def\div{\@ifstar{\@divergence}{\divergence}}
\def\curl{\@ifstar{\@curl}{\rotor}}
\def\rot{\@ifstar{\@rotor}{\rotor}}
\def\cond{\@ifstar{\@conditionnumber}{\conditionnumber}}
\def\laplacian{\@ifstar{\@laplaceoperator}{\laplaceoperator}}
\DeclareMathOperator*{\@gradient}{\mathrm{grad}}
\DeclareMathOperator*{\@curl}{\mathrm{curl}}
\DeclareMathOperator*{\@rotor}{\mathrm{rot}}
\DeclareMathOperator*{\@divergence}{\mathrm{div}}
\DeclareMathOperator*{\@conditionnumber}{\mathrm{cond}}
\DeclareMathOperator*{\@laplaceoperator}{\nabla^{2}}
\makeatother

\DeclareMathSymbol{\conv}{\mathrel}{symbols}{"03}
\DeclareMathSymbol{\comp}{\mathbin}{symbols}{"0E}
\DeclareMathSymbol{\grass}{\mathbin}{symbols}{"0E}

\DeclareMathSymbol{\mdot}{\mathord}{symbols}{"01} 
\DeclareMathSymbol{\mtimes}{\mathord}{symbols}{"02} 

\usepackage{wrapfig2}

\nocite{*}

\Russian

\usepackage{tabularx}

\renewcommand{\sectionFontShape}{\bfseries}
\renewcommand{\subsectionFontShape}{\bfseries}
\renewcommand{\subsubsectionFontShape}{\bfseries}
\renewcommand{\paragraphFontShape}{\bfseries}
\renewcommand{\subparagraphFontShape}{\bfseries}
\makeatletter
\def\@captionfont{\normalfont}
\makeatother
\usepackage{afterpage}
\usepackage[format=plain,labelsep=period,labelformat=simple,textformat=period,hypcap=true,]{subcaption}
\DeclareCaptionSubType*[asbuk]{figure}
\addto\extrasrussian{}
\addto\extrasenglish{}
\makeatletter
\renewcommand{\p@subfigure}{\thefigure{,}}
\makeatother
\DeclareCaptionLabelFormat{closing}{#2)}
\DeclareCaptionTextFormat{boldmath}{ \boldmath #1}
\DeclareRobustCommand{\typesetter}[1]%
{%
	\def\PFUtypesetter{#1}
}

\def\letUpgreekdefault{1} 
\def\letupgreekdefault{0} 

\usepackage{gost7184}
\AtBeginDocument%
{%
	\setdefaultleftmargin{2\parindent}{}{}{}{}{}%
	\selectlanguage{russian}%
	\Russian
}

\iflanguage{english}%
{%
	
	\edef\PFUaltlanguage{russian}
}%
{%
	
	\edef\PFUaltlanguage{english}
}

\makeatletter
\DeclareRobustCommand{\@biblabel}[1]{#1.\hfill}%
\makeatother

\makeatletter
\DeclareRobustCommand\factorial{\@ifnextchar[{\f@ctori@l}{\factori@l}}
\def\f@ctori@l[#1]#2{%
	\edef\@temp{#1}
	\edef\factorial@option{old}
	\ifx\@temp\factorial@option
		\mathpalette\f@ctori@l@ux{#2}
	\else
		\factori@l{#2}
	\fi
}
\def\factori@l#1{\mathpalette\factori@l@ux{#1}}
\def\f@ctori@l@ux#1#2{%
	{#1\mkern1mu\oalign{\vrule\,$#1#2\mathstrut$\,\cr\noalign{\hrule}}}}
\def\factori@l@ux#1#2{%
	{\oalign{$#1#2!\mathstrut$}}}
\makeatother

\makeatletter
\def\circled{\@ifnextchar[{\@circled}{\@circled[0.5pt]}}
\def\@circled[#1]{\@ifnextchar[{\@circled@opt[#1]}{\@circled@opt[#1][2pt]}}
\def\@circled@opt[#1][#2]#3{%
	\def\tempa{#1}%
	\def\tempb{#2}%
	\def\tempc{}%
	\ifx\tempa\tempc
		\def\tempa{0.5pt}
	\else\fi
	\ifx\tempb\tempc
		\def\tempb{2pt}
	\else\fi
	\tikz[baseline=(char.base)]{%
		\node[shape=circle,draw,inner sep=\tempb,line width=\tempa] (char) {#3};}}
\makeatother

\makeatletter
\def\ellipsed{\@ifnextchar[{\@ellipsed}{\@ellipsed[0.5pt]}}
\def\@ellipsed[#1]{\@ifnextchar[{\@ellipsed@opt[#1]}{\@ellipsed@opt[#1][2pt]}}
\def\@ellipsed@opt[#1][#2]#3{%
	\def\tempa{#1}%
	\def\tempb{#2}%
	\def\tempc{}%
	\ifx\tempa\tempc
		\def\tempa{0.5pt}
	\else\fi
	\ifx\tempb\tempc
		\def\tempb{2pt}
	\else\fi
	\tikz[baseline=(char.base)]{%
		\node[shape=ellipse,draw,inner sep=\tempb,line width=\tempa] (text) {#3};}}
\makeatother

\AtBeginDocument{%
	\addto\extrasrussian{%

	}
	\addto\extrasenglish{%

	}
}

\def\reportmonth#1{
	\ifcase#1\relax
		\def\RPRTmonth{Нулевой}
		\def\RPRTmonthshort{Нуль}
	\or
		\def\RPRTmonth{Январь}
		\def\RPRTmonthshort{Янв}
	\or
		\def\RPRTmonth{Февраль}
		\def\RPRTmonthshort{Фев}
	\or
		\def\RPRTmonth{Март}
		\def\RPRTmonthshort{Мар}
	\or
		\def\RPRTmonth{Апрель}
		\def\RPRTmonthshort{Апр}
	\or
		\def\RPRTmonth{Май}
		\def\RPRTmonthshort{Май}
	\or
		\def\RPRTmonth{Июнь}
		\def\RPRTmonthshort{Июн}
	\or
		\def\RPRTmonth{Июль}
		\def\RPRTmonthshort{Июл}
	\or
		\def\RPRTmonth{Август}
		\def\RPRTmonthshort{Авг}
	\or
		\def\RPRTmonth{Сентябрь}
		\def\RPRTmonthshort{Сен}
	\or
		\def\RPRTmonth{Октябрь}
		\def\RPRTmonthshort{Окт}
	\or
		\def\RPRTmonth{Ноябрь}
		\def\RPRTmonthshort{Ноя}
	\or
		\def\RPRTmonth{Декабрь}
		\def\RPRTmonthshort{Дек}
	\else
		\def\RPRTmonth{\fixme{Ошибка}}
		\def\RPRTmonthshort{\fixme{Ошибка}}
	\fi
}

\def\reportaltmonth#1{
	\ifcase#1\relax
		\def\RPRTaltmonth{Null}
		\def\RPRTaltmonthshort{Null}
	\or
		\def\RPRTaltmonth{January}
		\def\RPRTaltmonthshort{Jan}
	\or
		\def\RPRTaltmonth{February}
		\def\RPRTaltmonthshort{Feb}
	\or
		\def\RPRTaltmonth{March}
		\def\RPRTaltmonthshort{Mar}
	\or
		\def\RPRTaltmonth{April}
		\def\RPRTaltmonthshort{Apr}
	\or
		\def\RPRTaltmonth{May}
		\def\RPRTaltmonthshort{May}
	\or
		\def\RPRTaltmonth{June}
		\def\RPRTaltmonthshort{Jun}
	\or
		\def\RPRTaltmonth{July}
		\def\RPRTaltmonthshort{Jul}
	\or
		\def\RPRTaltmonth{August}
		\def\RPRTaltmonthshort{Aug}
	\or
		\def\RPRTaltmonth{September}
		\def\RPRTaltmonthshort{Sep}
	\or
		\def\RPRTaltmonth{October}
		\def\RPRTaltmonthshort{Oct}
	\or
		\def\RPRTaltmonth{November}
		\def\RPRTaltmonthshort{Nov}
	\or
		\def\RPRTaltmonth{December}
		\def\RPRTaltmonthshort{Dec}
	\else
		\def\RPRTaltmonth{\fixme{Error}}
		\def\RPRTaltmonthshort{\fixme{Error}}
	\fi
}

\def\pdftitle{The quantum mechanics of Einstein photons and generalized functions}
\def\pdfauthor{Beilinson A.\,A.}
\def\pdfkeywords{momentum and coordinate representations of generalized functions, Parseval equality, measure continuation, compact spaces.}
\def\pdfdisplaydoctitleprint{\pdftitle}

\def\pdfsubject{math-ph, quant-pm}

\typesetter{N.\,Ghonim}

\AtEndPreamble{%
	\hypersetup{%
		breaklinks=true,
		pdftitle=\pdftitle,
		pdfsubject=\pdfsubject,
		pdfauthor=\pdfauthor,
		pdfkeywords=\pdfkeywords,
		unicode=true,
		colorlinks=true, 
		anchorcolor=darkgreen,
		linkcolor=darkred,
		filecolor=darkteal,
		citecolor=darkgoldenrod,
		urlcolor=darkmagenta,
		pdfinfo={%
			pdftitle=\pdftitle,
			pdfsubject=\pdfsubject,
			pdfauthor=\pdfauthor,
			pdfkeywords=\pdfkeywords,
			pdfdisplaydoctitle=\pdfdisplaydoctitleprint,
			typesetter=\PFUtypesetter,
		},
	}
} 

\AtBeginDocument{%
	\addeditor[GN]{gn}{darkred}{darkcerulean}{vermilion}{darkyellow}{canaryyellow}{darkcerulean}%
}

\begin{document}
{\selectlanguage{english}



\title{%
	The quantum mechanics of Einstein photons\\
	and generalized functions%
}

\author[a]{Beilinson A.\,A.}

\address[a]{%
	Department of Theoretical Physics\\
	Russian Peoples' Friendship University named after Patrice Lumumba\\
	Russia, 117198, Moscow, Miklukho-Maklaya st., 6%
}

\email[a]{alalbeyl@gmail.com}
\orcid[a]{0000-0003-4474-9016}

\thanks{The author is grateful to Prof. R.\,A.\,Minlos for consultations and precious communication, to Prof. O.\,G.\,Smolyanov and Dr. N.\,Ringo for the support, and also to N.\,Ghonim for translation into English and \TeX nical assistance.}

\begin{abstract}
	The article consider an interpretation of Majorana equations as a quantum Lorentz covariant equations for the field of Einstein photon.

	A photon with ``deinterlaced'' spins (with diagonal Hamiltonian) is considered, its generalized Green function as a functional on finite test functions and its Schr\"{o}dinger equation are constructed. The generalized process corresponding to this Green function is continued to $\sigma$\=/additive quantum generalized measure on the space dual to the compact subspace of a photon paths in $L_{2}(-\infty,\infty)$. In this case, the filling of whole space by such quantum field instantly occurs in the evolutionary problem.

	The last part of the article gives the calculation of the ``Casimir forces'' arising in this field.
\end{abstract}


\date{}


\renewcommand{\headrulewidth}{0pt}
\fancyheadoffset{0pt}
\lhead{}
\rhead{}
\chead{}
\lfoot{}
\rfoot{}
\cfoot{\thepage}
\copyrightinfo{}{}

\maketitle







\section*{Introduction}

The article is based on an understanding of Majorana equations as a Lorentz covariant quan\-tum Schr\"{o}dinger equations for Einstein photon similar to Dirac equations for an electron. The generalized Green function as the functional on complex finite test (bump) functions is correspond to these equations.

The second part of work studies the quantum field of a photon devoided of its spin component interaction (a photon filed with ``deinterlaced'' spins), similar to Dirac field in Foldy and Wouthyusen variables. This field is interpreted as a quantum field the compact in $L_{2}(-\infty,\infty)$ with the generalized Green function as the functional on complex finite test functions, with Schr\"{o}dinger equation having no velocity operator.

Considering this generalized Green function as defining a generalized quantum process, we construct a continuation of this process to a generalized $\sigma$\=/additive quantum measure on the space dual to the space of photon trajectories with ``deinterlaced'' spins.

The third part calculates the quantity of ``Casimir forces'' as macroscopic forces appearing in the quantum field of a photon with ``deinterlaced'' spins.

The work uses notations and terminology accepted in monographs~\cite{gelfand:books:02:iss:01,gelfand:books:02:iss:04}. The velocity of light and Planck constant accepted equalled~$\num{1}$.

\section{Majorana variables and quantum theory of the e.-m. field}

It is known that from Maxwell equations
\begin{equation*}
	\pdv{E_{t}(x)}{t} = \rot*{H_{t}(x)}, \quad \pdv{H_{t}(x)}{t} = -\rot*{E_{t}(x)}
\end{equation*}
for e.-m. field outside of sources at the same time the wave equations
\begin{equation*}
	\pdv[ord={2}]{E_{t}(x)}{t} = \laplacian{E_{t}(x)}, \quad \pdv[ord={2}]{H_{t}(x)}{t} = \laplacian{H_{t}(x)},
\end{equation*}
follows for intensities $E_{t}(x)$ and $H_{t}(x)$.

Therefore, Majorana variables of e.-m. field
\begin{equation*}
	\mu_{t}(x) = E_{t}(x) + iH_{t}(x), \quad \overline{\mu_{t}(x)} = E_{t} - iH_{t}(x),
\end{equation*}
also satisfy the wave equations
\begin{equation*}
	\pdv[ord={2}]{\mu_{t}(x)}{t} = \laplacian{\mu_{t}(x)}, \quad \pdv[ord={2}]{\overline{\mu_{t}(x)}}{t} = \laplacian{\overline{\mu_{t}(x)}}.
\end{equation*}
As shown by Majorana~(see~\cite{axiezer:book:1981:qe}), the equations
\begin{equation*}
	\pdv{\mu_{t}(x)}{t} = -(S,\nabla) \mu_{t}(x),\quad
	\pdv{\mu_{t}^{+}(x)}{t} = -\mu_{t}^{+}(x) (S,\cev{\nabla}),
\end{equation*}
are true, where $\mu_{t}(x)$ is a column, $\mu_{t}^{+}(x)$ is a row, and \mbox{$S = iA$}, $A$ is infinitesimal rotation operators around coordinate axes in representation corresponding to weight~$\num{1}$, see~\cite{gelfand:book:1972:rglgp}:
\begin{equation*}
	S_{1} =
	\begin{pmatrix*}[r]
		0,& 0,& 0\\
		0,& 0,& -i\\
		0,& i,& 0\\
	\end{pmatrix*},\quad
	S_{2} =
	\begin{pmatrix*}[r]
		0,& 0,& i\\
		0,& 0,& 0\\
		-i,& 0,& 0\\
	\end{pmatrix*},\quad
	S_{3} =
	\begin{pmatrix*}[r]
		0,& -i,& 0\\
		i,& 0,& 0\\
		0,& 0,& 0\\
	\end{pmatrix*}.
\end{equation*}

Remark, that Majorana equations have a specific covariant form due to the complexity of the state vector $E + iH$ (see~\cite[p.~93]{landau:books:01:vol:02}), reducing the Lorentz transform to the rotation of this vector by an imaginary angle around an axis in $4$\=/dimensional space-time corresponding to this Lorentz transform.

Remark also, that, if we take one of Majorana equations\footnote{Ettore Majorana work was written (not published) soon after known work of Dirac on quantization of an electron and therefore the unity of approach when solving problem on e.-m. field quantization is traced. This is remarked by my colleague and disciple N.\,Ghonim.}
\begin{equation*}
	I\pdv*{%
		\begin{pmatrix}
			\mu_{t}^{(1)}(z)\\
			\mu_{t}^{(2)}(z)
		\end{pmatrix}}{t} = -%
		\begin{pmatrix*}[r]
			0,& -i\\
			i,& 0
		\end{pmatrix*} \pdv*{%
		\begin{pmatrix}
			\mu_{t}^{(1)}(z)\\
			\mu_{t}^{(2)}(z)
		\end{pmatrix}}{z},\quad\text{or shortly}\quad I\pdv*{}{t} = -%
		\begin{pmatrix*}[r]
			0,& -i\\
			i,& 0
		\end{pmatrix*} \pdv*{}{z},
\end{equation*}
and square this equation, then we obtain a wave equation \mbox{$I\pdv*[ord={2}]{}{t} = I\pdv*[ord={2}]{}{z}$} --- cf. with the procedure for extracting a matrix-valued root in Dirac equation theory.

In momentum representation the Majorana equations are become
\begin{equation*}
	i\pdv*{\tilde{\mu}_{t}(p)}{t} = -(S,p) \tilde{\mu}_{t}(p),\quad
	i\pdv*{\tilde{\mu}_{t}^{+}(p)}{t} = \tilde{\mu}_{t}^{+}(p) (S,p).
\end{equation*}

Let us multiply both parts of Majorana equa\-tions by a Planck constant $\hbar$:
\begin{equation*}
	\hbar\pdv*{\mu_{t}(x)}{t} = -\hbar(S,\nabla) \mu_{t}(x),\quad
	\hbar\pdv*{\mu_{t}^{+}(x)}{t} = -\hbar\mu_{t}^{+}(x) (S,\cev{\nabla}).
\end{equation*}
Remark, that these equations can be interpreted as Schr\"{o}dinger equations for a photon, especially that the spin of the photon is equal to $\hbar$.

Remark also, that these equations exist only with $\hbar \neq 0$; when $\hbar \to 0$ they disappear.

\noindent We will assume that the system of Majorana equations is the system of Schr\"{o}dinger equations for a photon.

Consider these Majorana--Schr\"{o}dinger equa\-tions for quantum field of a photon in more detail. To simplify record we will consider the one-dimensional quantum e.-m. field of a photon, which depends only on $z$, with the system of equations in momentum representation
\begin{equation*}
	i\pdv*{\tilde{\mu}_{t}(p_{z})}{t} = -S_{z} p_{z} \tilde{\mu}_{t}(p_{z}),\quad
	i\pdv*{\overline{\tilde{\mu}_{t}(p_{z})}}{t} = -S_{z} p_{z} \overline{\tilde{\mu}_{t}(p_{z})}
\end{equation*}
with the Green function \mbox{$\tilde{M}_{t}(p_{z}) = \exp\ab(itS_{z} p_{z})$}. Remark, that is a real exponential.

Bearing in mind the constructing of coordinate representation for the Green function $M_{t}(z)$ as a functional on columns of finite test functions $\varphi = \ab(\!\begin{smallmatrix*}[l]\varphi_{1}(z)\\\varphi_{2}(z)\end{smallmatrix*}\!)$, we will suppose $\tilde{M}_{t}(p_{z})$ is an analytic functional on columns of analytic test functions $\vartheta = \ab(\!\begin{smallmatrix*}[l]\vartheta_{1}(z)\\\vartheta_{2}(z)\end{smallmatrix*}\!)$,~see~\cite{gelfand:books:02:iss:01}.

Since, it is easy to see, we have (see the begin\-ning of second section for details of calculations)
\begin{equation*}
	\tilde{M}_{t}(p_{z}) =
	\begin{pmatrix*}[r]
		\cos tp_{z},& -\sin tp_{z}\\
		\sin tp_{z},& \cos tp_{z}
	\end{pmatrix*}
\end{equation*}
the orthogonal matrix and using Parseval equality
\begin{equation*}
	\int \overline{M_{t}}
	\begin{pmatrix*}[l]
		\varphi_{1}\\
		\varphi_{2}
	\end{pmatrix*}
	\odif{z} = \frac{1}{2\pi} \int \overline{\tilde{M}_{t}}
	\begin{pmatrix*}[l]
		\vartheta_{1}\\
		\vartheta_{2}
	\end{pmatrix*}
	\odif{p_{z}}
\end{equation*}
and Paley--Wiener theorem (see~\cite{gelfand:books:02:iss:01}), we obtain the generalized Green function
\begin{gather*}
	M_{t}(z) = \frac{1}{2}
	\begin{pmatrix*}[r]
		\delta_{_{+}}(z),& -i\delta_{_{-}}(z)\\
		i\delta_{_{-}}(z),& \delta_{_{+}}(z)
	\end{pmatrix*},\quad\text{where}\quad
		\delta_{_{+}}(z) = \delta(t - z) + \delta(t + z),\quad
		\delta_{_{-}}(z) = \delta(t - z) - \delta(t + z),
\end{gather*}
for quantum field of a photon in coordinate representation as a functional on columns of finite test functions, which is interpreted as two related waves divergent along axis $z$.

Let us assume now that columns of finite test functions are complex, namely
\begin{equation*}
	\begin{pmatrix*}[l]
		\psi_{1}(z)\\
		\psi_{2}(z)
	\end{pmatrix*} =
	\begin{pmatrix*}[l]
		\varphi_{1}(z)\\
		\varphi_{2}(z)
	\end{pmatrix*} + i
	\begin{pmatrix*}[l]
		\phi_{1}(z)\\
		\phi_{2}(z)
	\end{pmatrix*},
\end{equation*}
where $\varphi_{j}$ and $\phi_{j}$ $(j = 1,\,2)$ are real finite test functions. The columns can now be interpreted as a wave functions of a photon.

Indeed, since when $t$ is small the generalized Green function $M_{t}(z)$ becomes, it is easy to see,
\begin{equation*}
	\begin{pmatrix*}[r]
		\delta(z),& -it\delta'(z)\\
		it\delta'(z),& \delta(z)
	\end{pmatrix*},
\end{equation*}
we have
\begin{equation*}
	\begin{pmatrix*}[l]
		\psi_{t}^{(1)}(z)\\
		\psi_{t}^{(2)}(z)
	\end{pmatrix*} = \!\int
	\begin{pmatrix*}[r]
		\delta(z - \alpha),&\hspace*{-0.5em} -it\delta'(z - \alpha)\\
		it\delta'(z - \alpha),& \delta(z - \alpha)
	\end{pmatrix*}
	\begin{pmatrix*}[l]
		\psi_{0}^{(1)}(\alpha)\\
		\psi_{0}^{(2)}(\alpha)
	\end{pmatrix*} \odif{\alpha},
\end{equation*}
with the implication that when \mbox{$t \to 0$} the Schr\"{o}dinger equation
\begin{equation*}
	\pdv*{%
		\begin{pmatrix*}[l]
			\psi_{t}^{(1)}(z)\\
			\psi_{t}^{(2)}(z)
		\end{pmatrix*}}{t} =
	\begin{pmatrix*}[r]
		0,& -i\pdv{}{z}\\
		i\pdv{}{z},& 0
	\end{pmatrix*}
	\begin{pmatrix*}[l]
		\psi_{t}^{(1)}(z)\\
		\psi_{t}^{(2)}(z)
	\end{pmatrix*}
\end{equation*}
arises.

Consider this matrix equation (written by Majorana) with the Hermite Hamilton operator
\begin{equation*}
	\hat{H}_{z} =
	\begin{pmatrix*}[r]
		0,& -i\pdv{}{z}\\
		i\pdv{}{z},& 0
	\end{pmatrix*},
\end{equation*}
having a continuous trigonometric spectrum
\begin{equation*}
	\hat{H}_{z}
	\begin{pmatrix*}
		\cos(iEz)\\
		\sin(-iEz)
	\end{pmatrix*} = E
	\begin{pmatrix*}
		\cos(iEz)\\
		\sin(-iEz)
	\end{pmatrix*}.
\end{equation*}

Remark, that a wave functions for a photon belong to a Hilbert space that arises as an extension of the space of complex finite test columns $\ab(\!\begin{smallmatrix*}[l]\psi_{1}(z)\\\psi_{2}(z)\end{smallmatrix*}\!)$ with the square of the Hilbert length
\begin{equation*}
	\ab(%
		\begin{pmatrix*}[l]
			\psi_{1}(z)\\
			\psi_{2}(z)
		\end{pmatrix*}^{+}
		\begin{pmatrix*}[l]
			\psi_{1}(z)\\
			\psi_{2}(z)
		\end{pmatrix*}%
	) = \int \ab(\overline{\psi_{1}(z)},\,\overline{\psi_{2}(z)})
		\begin{pmatrix*}[l]
			\psi_{1}(z)\\
			\psi_{2}(z)
		\end{pmatrix*} \odif{z} = \int \ab(\abs{\psi_{1}(z)}^{2} + \abs{\psi_{2}(z)}^{2}) \odif{z}.
\end{equation*}

Remark also, that the position probability density of a photon in a state
\begin{equation*}
	\ab(\!\begin{smallmatrix*}[l]\psi_{t}^{(1)}(z)\\\psi_{t}^{(2)}(z)\end{smallmatrix*}\!) \quad\text{is}\quad \frac{\abs{\psi_{t}^{(1)}(z)}^{2} + \abs{\psi_{t}^{(2)}(z)}^{2}}{\int \ab(\abs{\psi_{t}^{(1)}(z)}^{2} + \abs{\psi_{t}^{(2)}(z)}^{2}) \odif{z}},
\end{equation*}
which, it is easy to see, coincides with the normalized energy density of classic e.-m. field (cf.~with the end of third section of this article).

Let us construct the velocity operator $\hat{V}_{z}$ for a photon field as the commutator of the coordinate operator and Hamiltonian. It is easy to see, that
\begin{equation*}
	\hat{V}_{z} = \comm[\bigg]{\hat{H}_{z}}{%
		\begin{pmatrix*}[r]
				z,& 0\\
				0,& z
		\end{pmatrix*}} =
	\begin{pmatrix*}[r]
		1,& 0\\
		0,& 1
	\end{pmatrix*}.
\end{equation*}
Therefore $\hat{V}_{z}$ is Hermite unitary operator in Hilbert space of a photon states with eigenvalues $\num{\pm 1}$.

And therefore the average values of the operator $\hat{V}_{z}$ on a photon states turn out to be equal to either one or $\num{-1}$, which corresponding to the structure noted above of the Green function for quantum field of a photon as two related waves, divergent with fundamental velocity along axis $z$.

But this implies that the photon quantum field propagation velocity measured with help of instruments is always equal to the velocity of light.

At the same time, since \mbox{$\hat{H}_{z} = \hat{V}_{z} \pdv*{}{z}$}, the Majorana--Schr\"{o}dinger equation for a photon field in the considered one-dimensional case is
\begin{equation*}
	\pdv*{%
		\begin{pmatrix*}[l]
			\psi_{t}^{(1)}(z)\\
			\psi_{t}^{(2)}(z)
		\end{pmatrix*}}{t} + \hat{V}_{z} \pdv*{%
		\begin{pmatrix*}[l]
			\psi_{t}^{(1)}(z)\\
			\psi_{t}^{(2)}(z)
		\end{pmatrix*}}{z} = 0,
\end{equation*}
and, it is easy to see, that in the spatial case
\begin{equation*}
	\pdv*{\mu_{t}(x)}{t} + \scalar{\hat{V}}{\nabla} \mu_{t}(x) = 0.
\end{equation*}

\begin{statement*}
	The Majorana variables can be considered both as equations of the classic e.-m. field and as the quantum covariant equations of the photon field. Here the field of a photon ``considered'' with help of classic instruments is ``seen'' as a classic field.
\end{statement*}

\section{Generalized quantum Cauchy measure of photon paths\\in Einstein photon theory}

Let us consider disclosure of the exponential
\begin{equation*}
	\tilde{M}_{t}(p_{z}) = \exp\ab(itS_{z} p_{z})
\end{equation*}
in more detail.

Since the matrix $S_{z} p_{z}$ in the exponent is Hermitian, it can be reduced to diagonal form by the following unitary transform:
\begin{equation*}
	\tilde{M}_{t}(p_{z}) = \tilde{Q}^{+}(p_{z}) \exp
	\begin{pmatrix*}[r]
		it\abs{p_{z}},& 0\\
		0,& -it\abs{p_{z}}
	\end{pmatrix*} \tilde{Q}(p_{z}) =
	\begin{pmatrix*}[r]
		\cos tp_{z},& -\sin tp_{z}\\
		\sin tp_{z},& \cos tp_{z}
	\end{pmatrix*},
\end{equation*}
where
\begin{equation*}
	\tilde{Q}(p_{z}) = \frac{1}{\sqrt{2}}
	\begin{pmatrix*}[r]
		i\sgn p_{z},& 1\\
		-i\sgn p_{z},& 1
	\end{pmatrix*}.
\end{equation*}
The subject of consideration will be the resulting quantum field (Cauchy field), which is naturally understood as the quantum field of a photon with ``deinterlaced'' spins with the generalized Green function in the momentum representation
\begin{equation*}
	\exp
	\begin{pmatrix*}[r]
		it\abs{p_{z}},& 0\\
		0,& -it\abs{p_{z}}
	\end{pmatrix*},
\end{equation*}
that is unitarily equivalent to the field of an usual photon.

Keep in mind that the electron-positron Dirac field in the Foldy--Wouthyusen variables~\cite{foldy:article:1950:dt} is turn out to be almost the same quantum Cauchy field, as it will be published later.

We will use our previously constructed coordi\-nate representation of the generalized Green func\-tion of this one-dimensional quantum field as the quantum Cauchy field (see~\cite{bejlinson:article:2018:02})
\begin{equation*}
	C_{it}(z) = \frac{1}{2} \ab\big(\delta(t - z) + \delta(t + z)) + \frac{i}{\pi} \ab(\frac{t}{t^{2} - z^{2}})
\end{equation*}
as a functional on finite test complex functions $\Psi(z) = \Phi(z) + i\mathrm{X}(z)$.

Let us take advantage of the fact that the $\int \overline{C_{it}(z)} \Psi(z) \odif{z}$ is an even generalized Green function and construct the Schr\"{o}dinger equation corresponding to it. For this purpose, consider it when $t$ are small.

Remark, that the functional
\begin{equation*}
	\int \frac{1}{2} \ab(\frac{1}{t - z} + \frac{1}{t + z}) \Psi(z) \odif{z} = t\int \frac{1}{2t} \ab(\frac{1}{t - z} - \frac{1}{-t - z}) \Psi(z) \odif{z}
\end{equation*}
\noindent is
\begin{equation*}
	t\int \pdv*[delims-eval={.}{|}]{\frac{\Psi(z)}{t - z}}{t}_{t = 0} \odif{z} = -t\int \ab(\pdv*{\frac{1}{z}}{z}) \Psi(z) \odif{z} = t\int \frac{\Psi'(z)}{z} \odif{z},
\end{equation*}
when $t$ are small. Here the integral should be understood in the sense of Cauchy principal value (see~\cite[p.~52:~formula~(8)]{gelfand:books:02:iss:01}), that is why it should be assumed
\begin{equation*}
	\int \frac{\Psi'(z)}{z} \odif{z} = \lim_{\varepsilon \to 0} \ab(\,\int_{-\infty}^{-\varepsilon} + \int_{\varepsilon}^{\infty}\,) \frac{\Psi'(z)}{z} \odif{z},
\end{equation*}
where the generalized function $\frac{1}{z}$ has discontinuities in points $-\varepsilon$ and $\varepsilon$ (see~\cite[p.~22:~formula~(1)]{gelfand:books:02:iss:01}), and we will obtain
\begin{equation*}
	\lim_{\varepsilon \to 0} \ab(\,\int_{-\infty}^{-\varepsilon} + \int_{\varepsilon}^{\infty}\,) \frac{\Psi'(z)}{z} \odif{z} = \lim_{\varepsilon \to 0} \int \ab(\frac{\delta(z + \varepsilon)}{-\varepsilon} + \frac{\delta(z - \varepsilon)}{\varepsilon}) \Psi(z) \odif{z} = 2\int \delta'(z) \Psi(z) \odif{z}.
\end{equation*}

Therefore, recalling that $C_{it}(z)$ is generalized Green function, when $t$ are small we have
\begin{equation*}
	\Psi_{t}(z) = \Psi_{0}(z) + \frac{2it}{\pi} \int \delta(z - \alpha) \Psi_{0}(\alpha) \odif{\alpha},
\end{equation*}
and, when $t \to 0$, the Cauchy--Schr\"{o}dinger equation at the initial, and therefore at any moment of time
\begin{equation*}
	\pdv*{\Psi_{t}(z)}{t} = -\frac{2i}{\pi} \pdv*{\Psi_{t}}{z}
\end{equation*}
with Hamiltonian \mbox{$\hat{h} = -\frac{2i}{\pi} \pdv*{}{z}$} in the space of complex finite test functions $\Psi(z)$.

Keep in mind that the velocity operator \mbox{$\hat{V} = \comm{-\frac{2i}{\pi} \pdv*{}{z}}{z} = ic$} ($c$ is real number) of this quan\-tum field is not Hermitian and the velocity in the states of such quantum field is fundamentally not determined. Let us not forget that \mbox{$\Psi(z) = \Phi(z) + i\mathrm{X}(z)$}, where $\Phi(z)$, $\mathrm{X}(z)$ are even finite test functions.

We will find the totality of Hilbert functions, on which the Hamiltonian \mbox{$\hat{h} = -\frac{2i}{\pi} \pdv*{}{z}$} is defined, as an extension of the space of complex finite test functions~$\Psi_{t}(z)$.

Let us take advantage of the fact that Cauchy functional is a generalized Green function and therefore defines a Cauchy process~(see~\cite{gelfand:books:02:iss:04}).

Note, that on each partition \mbox{$\adif{t_{j}} = t_{j} - t_{j - 1}$} \mbox{$(j = 1, \ldots, n)$} of the time segment\footnote{We will assume this segment of time and its partition are fixed until this restriction is lifted.} $\brange{0,t}$ the Cauchy functional is defined on cylindric sets of trajectories~(see~\cite[Ch.~4:~\S1]{gelfand:books:02:iss:04}).

To constructing a Cauchy functional on a cylindric set of trajectories $a = \sum_{j = 1}^{n} \alpha_{j} z_{j}$, it is easy to see, we should consider
\begin{equation*}
	\int \ab\Big(\prod_{j = 1}^{n} \overline{C_{i\!\adif{t_{j}}}(z_{j})}) \Psi\ab\Big(\sum_{j = 1}^{n} \alpha_{j} z_{j}) \odif{z_{1}} \ldots \odif{z_{n}}.
\end{equation*}
Since \mbox{$\Psi(z) = \frac{1}{2\pi} \int \exp(-ipz) \Lambda(p) \odif{p}$}, where $\Lambda(p) = \Xi(p) + i\Upsilon(p)$, $\Xi(p)$ and $\Upsilon(p)$ are analytic test functions (Paley--Wiener theorem), we have
\begin{multline*}
	\hspace*{-1em}\int \ab\Big(\prod_{j = 1}^{n} \overline{C_{i\!\adif{t_{j}}}(z_{j})}) \Psi\ab\Big(\sum_{j = 1}^{n} \alpha_{j} z_{j}) \prod_{j = 1}^{n} \odif{z_{j}} = \frac{1}{2\pi} \int \prod_{j = 1}^{n} \int \overline{C_{i\!\adif{t_{j}}}(z_{j})} e^{-ip\alpha_{j} z_{j}} \odif{z_{j}} \Lambda(p) \odif{p} =\\
	= \frac{1}{2\pi} \int \prod_{j = 1}^{n} \exp\ab\big(-i\abs*{p\alpha_{j}} \adif{t_{j}}) \Lambda(p) \odif{p} = \frac{1}{2\pi} \int \overline{C_{iT}(r)} \Psi(r) \odif{r},
\end{multline*}
where $\Psi(r) = \Phi(r) + i\mathrm{X}(r)$, $T = \sum_{j = 1}^{n} \abs*{\alpha_{j}} \adif{t_{j}}$.

To constructing a Cauchy functional on a cylindric set of trajectories \mbox{$a_{k} = \sum_{j = 1}^{n} \alpha_{jk} z_{j}$} \mbox{$(k = 1, \ldots, n)$}, where $\alpha_{jk}$ is $n\mtimes n$ orthogonal matrix, consider the functional
\begin{equation*}
	\int \ab\Big(\prod_{j = 1}^{n} \overline{C_{i\!\adif{t_{j}}(z_{j})}}) \Psi\ab\Big(\sum_{j = 1}^{n} \alpha_{j1} z_{j},\ldots,\sum_{j = 1}^{n} \alpha_{jn} z_{j}) \prod_{j = 1}^{n} \odif{z_{j}}.
\end{equation*}
Literally repeating the constructing of the pre\-vious functional, we obtain the momentum repre\-sentation of desired functional
\begin{equation*}
	\int \overline{\tilde{C}_{i\!\adif{t}}(p)} \Lambda(p) \odif{p_{1}} \ldots \odif{p_{n}}
\end{equation*}
as an analytic functional on analytic test functions, where
\begin{equation*}
	\tilde{C}_{i\!\adif{t}}(p) = \exp\ab\Big(i\sum_{j = 1}^{n} \abs[\Big]{\sum_{k = 1}^{n} \alpha_{jk} p_{k}} \adif{t_{j}}),\quad \adif{t} = \adif{t_{1}},\ldots,\adif{t_{n}},\quad p = p_{1},\ldots,p_{n}.
\end{equation*}

The coordinate representation
\begin{equation*}
	C_{i(\adif{t_{1}},\ldots,\adif{t_{n}})}(a_{1},\ldots,a_{n}) = C_{i\!\adif{t}}(a)
\end{equation*}
of the desired functional on complex finite test functions is given by the Parseval equality
\begin{equation*}
	\int \overline{C_{i\!\adif{t}}(a)} \Psi(a) \odif{a} = \frac{1}{(2\pi)^{n}} \int \overline{\tilde{C}_{i\!\adif{t}}(p)} \Lambda(p) \odif{p},
\end{equation*}
which is the result of the direct construction of the functional $\int \overline{C_{i\!\adif{t}}(a)} \Psi(a) \odif{a}$ by a multilinear functional $\prod_{j = 1}^{n} \int \overline{C_{i\!\adif{t_{j}}}(r)} \Psi(r) \odif{r}$.

Thus the Cauchy functional on the cylindric sets of trajectories $a_{k} = \sum_{j = 1}^{n} \alpha_{jk} z_{j}$ $(k = 1, \ldots, n)$ is constructed.

Let us take advantage of the fact that the Cauchy functional is defined on hyperplanes \mbox{$(a,\omega) = r_{\omega}$} in~$R^{(n)}$ with unit normal vectors $\omega$ $\bigl(${}$a$, $\omega$, $p$ are vectors in $R^{(n)}$, $\abs{\omega} = 1${}$\bigr)$.

Indeed, the Cauchy functional on each such cylindric set is
\begin{multline*}
	\int \overline{C_{i\!\adif{t}}(a)} \Psi\ab\big((a,\omega)) \odif{a} = \frac{1}{2\pi} \int \overline{C_{i\!\adif{t}}(a)} \exp\ab\big(-ip(a,\omega)) \Lambda(p) \odif{p,a} =\\
	= \frac{1}{2\pi} \int \exp\ab\bigg(i\abs{p} \sum_{j = 1}^{n} \abs*{\sum_{k = 1}^{n} \alpha_{jk} \omega_{k}} \adif{t_{j}}) \Lambda(p) \odif{p} = \int \overline{C_{iT_{\omega}^{(n)}}(r)} \Psi(r) \odif{r},
\end{multline*}
where $\Psi(r) = \Phi(r) + i\mathrm{X}(r)$, $T_{\omega}^{(n)} = \sum_{j = 1}^{n} \abs[\Big]{\sum_{k = 1}^{n} \alpha_{jk} \omega_{k}} \adif{t_{j}}$.

Consider $n$ vectors $\omega^{(1)},\,\ldots,\,\omega^{(n)}$; on each hyperplane \mbox{$r_{j} = (a,\omega^{(j)})$} the Cauchy functional~$\int \overline{C_{iT_{\omega^{(j)}}^{(n)}}(r)} \Psi(r) \odif{r}$ is defined.

We fix each vector $\omega^{(j)}$ by the condition of collinearity with the vector \mbox{$\alpha_{j} = (\alpha_{j1},\ldots,\alpha_{jn})$}; then we obtain $n$ Cauchy functionals $\int \overline{C_{iT_{\omega^{(j)}}^{(n)}}(r)} \Psi(r) \odif{r}$, where
\begin{equation*}
	T_{\omega^{(j)}}^{(n)} = \sum_{j = 1}^{n} \abs[\Big]{\sum_{k = 1}^{n} \alpha_{jk} \omega_{k}^{(j)}} \adif{t_{j}} = \sum_{j = 1}^{n} \sqrt{\alpha_{j1}^{2} + \ldots + \alpha_{jn}^{2}} \adif{t_{j}}.
\end{equation*}

Therefore, the Cauchy functional
\begin{equation*}
	\int \overline{C_{i\sum_{j = 1}^{n} T_{\omega^{(j)}}^{(n)}}(r)} \Psi(r) \odif{r}
\end{equation*}
is defined.

Recall, that we are constructing the space extension of complex finite test functions, on which the Hamiltonian \mbox{$\hat{h} = -\frac{2i}{\pi} \pdv*{}{z}$} is defined. Let us show that this extension is the space, compact in~$L_{2}(-\infty,\infty)$.

Consider a sum $\sum_{j = 1}^{n} \sqrt{\alpha_{j1}^{2} + \ldots + \alpha_{jn}^{2}} \adif{t_{j}}$ when $n \to \infty$. Remark, that there is
\begin{equation*}
	\lim_{n \to \infty} \sum_{j = 1}^{n} \sqrt{\alpha_{j1}^{2} + \ldots + \alpha_{jn}^{2}} \adif{t_{j}} = \int_{0}^{t} \sqrt{\int \mathrm{A}_{\tau}^{2}(z) \odif{z}} \odif{\tau} = \int_{0}^{t} \norm{\mathrm{A}_{\tau}} \odif{\tau}
\end{equation*}
as a Riemann integral, which follows for an arbitrary partition of the segment $\brange{0,t}$ from the limitation of norms $\norm{\mathrm{A}_{\tau}}$.

Let us pay attention to the fact that here~$\tau$ can be considered as an index, that fixes a specific Hilbert function $\mathrm{A}_{\tau}(z)$. Therefore there is $\int \overline{C_{i\int_{0}^{t} \norm{\mathrm{A}_{\tau}} \odif{\tau}}(r)} \Psi(r) \odif{r}$.

Consider the consequences.

\noindent Recall, that up till now only one partition of the time segment $\brange{0,t}$ could be used; only since the last formula is obtained this restriction is removed (the quantity of the time segment $\brange{0,t}$ remains fixed). In this connection, a compact totality of vectors $\mathrm{A}_{\tau}(z)$ $(0 \leqslant \tau \leqslant t)$ arose --- since the totality $\mathrm{A}_{\tau}(z)$ is isomorphic to the set of points on segment $\brange{0,t}$, representing together the real compact Hilbert space~$L_{2}^{(c)}(-\infty,\infty)$. Remind, the vectors $\mathrm{A}_{\tau}(z)$ are unnormalized.

Let us use the Hilbert--Schmidt theorem~\cite{kolmogorov:book:1999:etffa}, according to which in $L_{2}^{(c)}(-\infty,\infty)$ there exists a discrete orthonormal basis $\xi_{k}(z)$ $(k = 1,\,2,\,\ldots,\,\infty)$ of the eigenvectors of the Hermite Hamiltonian $\hat{h}$ with eigenvalues $e_{k}$ such that $e_{k} \to 0$ when $k \to \infty$.

Therefore, there is defined the space, on which the Hamiltonian $\hat{h}$ is defined, as an extension of the space of complex finite test functions $\Psi(r)$ to a compact complex Hilbert space of functions $(0 \leqslant \tau \leqslant t)$
\begin{equation*}
	\psi_{\tau}^{(c)}(z) = \varphi_{\tau}^{(c)}(z) + i\phi_{\tau}^{(c)}(z),
\end{equation*}
with the Cauchy--Schr\"{o}dinger equation
\begin{equation*}
	\pdv*{\psi_{\tau}^{(c)}(z)}{\tau} + \frac{2i}{\pi} \pdv*{\psi_{\tau}^{(c)}(z)}{z} = 0
\end{equation*}
for this quantum Cauchy field.

Recall, that the Cauchy functional $\int{\overline{C_{iT_{\omega^{(j)}}^{(n)}}(r)} \Psi(r) \odif{r}}$ was constructed on $n$ cylindric sets of trajectories \mbox{$r = (a,\omega^{(j)})$} corresponding to $n\mtimes n$ orthogonal matrix, and as it now turns out, continued on the space of functionals on the complex compact space $L_{2}^{(c)}(-\infty,\infty)$. If we now understand the Cauchy functional on a cylindric sets of trajectories as generalized Cauchy measure (a Cauchy premeasure) of the totality of these tra\-jectories, then this premeasure becomes $\sigma$\=/additive generalized Cauchy measure on a functionals on $L_{2}^{(c)}(-\infty,\infty)$, see~\cite[ch.~4]{gelfand:books:02:iss:04}, since in $L_{2}^{(c)}(-\infty,\infty)$ the Cauchy functionals on cylindric sets belonging to the exterior of the sufficiently large radius ball are equal to zero. Therefore, the totality of trajectories on the segment $\brange{0,t}$, belonging to~$L_{2}^{(c)}(-\infty,\infty)$, is the support of the quantum Cauchy measure.

But this betokens the following.

\noindent Consider
\begin{equation*}
	\sqrt{\int \abs[\big]{\psi_{\tau}^{(c)}(z)}^{2} \odif{z}} = a(\tau),
\end{equation*}
where $\psi_{\tau}^{(c)}(z)$ is element of $\mathrm{A}_{\tau}(z)$. It is clear that $a(\tau)$ are positive equibounded and equicontinuous functions on the segment $\brange{0,t}$.

However, it is easy to see, the totality of $a(\tau)$ is equibounded and equicontinuous functions in $C(0,t)$. Therefore the totality of such functions $C^{(c)}(0,t)$ is compact in $C(0,t)$ (Arzela condition).

Therefore, there is defined a Cauchy functional
\begin{equation*}
	\int \overline{C_{i\int_{0}^{t} a(\tau) \odif{\tau}}(r)} \Psi(r) \odif{r}
\end{equation*}
on each function $a(\tau)$ (on ``Feynman paths'') as on the compact support of the Cauchy measure.

\begin{theorem*}
	Generalized quantum Cauchy process on complex finite test functions on the time segment~$\brange{0,t}$ is continued on the space $L_{2}^{(c)}(-\infty,\infty)$ of compact complex functions~$\psi_{\tau}^{(c)}(z)$, thereby becoming a generalized quantum Cauchy measure, which is $\sigma$\=/additive on this space. At the same time the quantum Cauchy measure is defined on the space of positive compact continuous functions~$C^{(c)}(0,t)$ related to $L_{2}^{(c)}(-\infty,\infty)$.
\end{theorem*}

\section{Casimir forces and quantum Cauchy field}

It is known, that the Casimir forces are called experimentally observed forces of attraction between metallic neutral mirrors, growing as $d^{-4}$ ($2d$ is distance between mirrors). We will assume that the time axis lies in a plane parallel to the mirrors in the middle between them, and the $z$ axis is orthogonal to the mirrors. We show that the Casimir forces arise in the quantum e.-m. field in states~$\psi_{\tau}^{(c)}(z) \in L_{2}^{(c)}(-\infty,\infty)$.

Recall, that solutions $\psi_{\tau}^{(c)}(z)$ \mbox{$(\tau \in \brange{0,t})$} of Cauchy--Schr\"{o}dinger equation are a complex compact functions
\begin{equation*}
	\psi_{\tau}^{(c)}(z) = \varphi_{\tau}^{(c)}(z) + i\phi_{\tau}^{(c)}(z),
\end{equation*}
the norms of which are a functionals, and the generalized quantum $\sigma$\=/additive Cauchy measure is defined on latter.

But this indicates that
\begin{equation*}
	\int \!\overline{C_{i\sum_{j} \norm{\psi_{j}^{(c)}}}(r)} \Psi(r) \odif{r} \!=\! \!\sum_{j}\! \int \!\overline{C_{i\norm{\psi_{j}^{(c)}}}(r)} \Psi_{j}(r) \odif{r}.
\end{equation*}

Therefore, the influence of the mirrors is reduced to excision (isolation) a part of the quantum e.-m. field lying between them as a separate quantum field in state \mbox{$\psi_{j\tau}^{(c)}(z) \in L_{2}^{(c)}(-\infty,\infty)$}.

This is inadmissibly for any quantum field in $L_{2}(-\infty,\infty)$ not belonging to $L_{2}^{(c)}(-\infty,\infty)$, which follows from the non-additivity of the energy of arbitrary quantum field in $L_{2}(-\infty,\infty)$.

We will find the energy of the entire quantum Cauchy field in the state
\begin{equation*}
	\psi_{\tau}^{(c)}(z) = \sum_{j = 1}^{\infty} c_{j} \exp(e_{j} \tau) \xi_{j}(z)
\end{equation*}
as the average of its Hamiltonian $\hat{h}$ over quantum field states involved in the formation of this field.
\begin{equation*}
	\overline{\hat{h}}^{\psi_{\tau}^{(c)}(z)} = \frac{\displaystyle\int \overline{\psi_{\tau}^{(c)}(z)} \hat{h} \psi_{\tau}^{(c)}(z) \odif{z}}{\displaystyle\int \overline{\psi_{\tau}^{(c)}(z)} \psi_{\tau}^{(c)}(z) \odif{z}} = \dfrac{\displaystyle\sum\nolimits_{j = 1}^{\infty} \abs{c_{j}}^{2} \exp(2\abs{e_{j}} \tau) \abs{e_{j}}}{\displaystyle\sum\nolimits_{j = 1}^{\infty} \abs{c_{j}}^{2} \exp(2\abs{e_{j}} \tau)}.
\end{equation*}

Therefore, the energy of the entire quantum Cauchy field is defined and finite. We will consider only even states of the field $\psi_{\tau}^{(c)}(0) \neq 0$, since only in such states of the field does the Casimir force arise.

Remark, that in even states the energy of half of the field $\frac{1}{2} \overline{\hat{h}}^{\psi_{\tau}^{(c)}(z)}$ is defined.

\noindent But, due to $\sigma$\=/additivity of the quantum Cauchy measure on functionals on $\psi_{\tau}^{(c)}(z) \in L_{2}^{(c)}(-\infty,\infty)$, the energy of any part of this quantum field, representable as $\bigcup_{j} \psi_{j\tau}^{(c)}(z)$, is defined. Therefore, the energy of the part of the field from $z = 0$ to $z = d$ (assuming the energy level $\abs{e_{k}}$ to be closest to $d$) is
\begin{equation*}
	E_{k} = \frac{\displaystyle\int\nolimits_{0}^{\infty} \sum\nolimits_{j = k}^{\infty} \overline{\psi_{j\tau}^{(c)}(z)} \abs{e_{j}} \psi_{j\tau}^{(c)}(z) \odif{z}}{\displaystyle\int\nolimits_{0}^{\infty} \sum\nolimits_{j = 1}^{\infty} \overline{\psi_{j\tau}^{(c)}(z)} \psi_{j\tau}^{(c)}(z) \odif{z}}.
\end{equation*}

Then the force, as a derivative of $E_{k}$ by $d$ for large $k = N$, is
\begin{equation*}
	F = \lim_{d \to 0} \frac{1}{d} \frac{\displaystyle\sum\nolimits_{j = N}^{\infty} \abs{c_{j}}^{2} \abs{e_{j}} \exp(2\abs{e_{j}} \tau)}{\displaystyle\sum\nolimits_{j = 1}^{\infty} \abs{c_{j}}^{2} \exp(2\abs{e_{j}} \tau)}.
\end{equation*}

Here the numerator of the fraction is the energy of a part of the field lying between $0$ and $d$, besides the energy of this field is discrete, and the energy level $e_{j}$ are such that \mbox{$e_{j} \to 0$} when \mbox{$j \to \infty$}. But, according to Planck, \mbox{$e_{j} = \omega$} (recall, that the quantity of Planck constant is taken to be equal \num{1}, and $\omega$ is a frequency of this field) and therefore, starting from a certain number $N$, the quantities $e_{j}$ turn out to be less than $\omega$, and corresponding to them the e.-m. field will not be able to be either radiated or absorbed. That is why, starting from this $N$ the numerator of the fraction ceases to be decreased when shifting the mirrors.

In addition, both the numerator and the denomi\-nator of the fraction remain non-zero, and the fraction itself is small. And the energy levels in such states of the photon are negative.

Remark, that the force $F$ can be considered as arising in the compact quantum field of a photon with ``deinterlaced'' spins, represented as a segment of $t$ length in the plane $z = 0$, as the ``Feynman path'' of such photon.

Remark also, that the constructed force $F$ corresponds to a segment of length $t$ on the time axis, while experimentally the force applied to a mirror area unit was measured. Therefore, taking into account the presence of the complex conjugate Cauchy field in the expression for direct product of the ``deinterlaced'' photon fields, we have the Casimir force
\begin{equation*}
	F_{\text{Casimir}} = \ab(\frac{F}{t})^{4}.
\end{equation*}

\section*{Conclusion}

The attribution of Majorana equations as the equations of a quantum field of a photon defined by the generalized Green function has been carried out. The procedure of ``deinterlacing'' the spin components of a photon led to the appearance of a quantum field with states compact in $L_{2}(-\infty,\infty)$, which does not have a velocity operator, with a $\sigma$\=/additive quantum Cauchy measure on the functionals on states in~$L_{2}^{(c)}(-\infty,\infty)$ or to states on elements $C^{(c)}(0,t)$ --- on ``Feynman paths'' of such photon.

However, it is in this quantum field that Casimir forces arise generating also well-known ``Lamb shift''~\cite{bejlinson:article:2024:01}.

Note, that during evolution the Cauchy field fills instantly any bounded part of the Euclidean space, which is discovered, apparently experimentally~\cite{nobelprize:physics:2022:01}, and optical instruments should not affect this quantum field at all.

\section*{Declarations}

\paragraph{Funding.} The authors did not receive support from any organization for the submitted work.

\paragraph{Conflict of interest.} The authors have no relevant financial or non-financial interests to disclose.

\paragraph{Data availability.} We do not analyse or gene\-rate any datasets, because our work proceeds within a theoretical and mathematical approach. One can obtain the relevant materials from the references below.

\putbib[default]


} 

\end{document}